\documentclass[a4paper,11pt]{article}
\pdfoutput=1 

\usepackage{jcappub} 

\usepackage{aas_macros}
\usepackage{natbib}
\usepackage{amsfonts}
\usepackage{amsmath}
\usepackage{amssymb}
\usepackage{graphicx}
\usepackage[dvipsnames]{xcolor}
\usepackage{hyperref}
\hypersetup{colorlinks=true,allcolors=teal}

\newcommand{\Mpch}{\ensuremath{h^{-1}{\rm Mpc}}}
\newcommand{\hMpc}{\ensuremath{h\,{\rm Mpc}^{-1}}}

\newcommand{\kms}{\ensuremath{{\rm km\,s}^{-1}}}

\newcommand{\e}[1]{\ensuremath{{\rm e}^{#1}}}

\newcommand{\erf}[1]{\ensuremath{{\rm erf}\left(#1\right)}}

\newcommand{\eqn}[1]{equation~\eqref{#1}}
\newcommand{\eqns}[1]{equations~\eqref{#1}}

\newcommand{\be}{\begin{equation}}
\newcommand{\ee}{\end{equation}}

\newcommand{\filgen}{\texttt{FilGen}}
\newcommand{\filapt}{\texttt{FilAPT}}

\usepackage{multirow}
\usepackage{tabularx}
\newcommand{\rsp}{\ensuremath{r_{\rm{sp}}}}

\newcommand{\disp}{DisPerSE}
\newcommand{\Nsm}{\ensuremath{N_{\rm{sm}}}}
\usepackage[normalem]{ulem}

\title{Towards unbiased recovery of cosmic filament properties: the role of spine curvature and optimized smoothing}

\author[1]{Saee Dhawalikar,\note{Corresponding author.}}
\author{and Aseem Paranjape}
\affiliation{Inter-University Centre for Astronomy \& Astrophysics,\\ Ganeshkhind, Post Bag 4, Pune 411007, India}

\emailAdd{saee.dhawalikar@iucaa.in}
\emailAdd{aseem@iucaa.in}

\abstract{
Cosmic filaments, the most prominent features of the cosmic web, possibly hold untapped potential for cosmological inference. 
While it is natural to expect the structure of filaments to show universality similar to that seen in dark matter halos, the lack of agreement between different filament finders on what constitutes a filament has hampered progress on this topic. 
We initiate a programme to systematically investigate and uncover possible universal features in the phase space structure of cosmic filaments, by generating particle realizations of mock filaments with \emph{a priori} known properties. 
Using these, we identify an important source of bias in the extraction of radial density profiles, which occurs when the local curvature $\kappa$ of the spine exceeds a threshold determined by the filament thickness. 
This bias exists even for perfectly determined spines, thus affecting \emph{all} filament finders. 
We show that this bias can be nearly eliminated by simply discarding the regions with the highest $\kappa$, with little loss of precision. 
An additional source of bias is the noise generated by the filament finder when identifying the spine, which depends on both the finder algorithm as well as intrinsic properties of the individual filament. 
We find that to mitigate this bias, it is essential not only to smooth the estimated spine, but to \emph{optimize} this smoothing separately for each filament. 
We propose a novel optimization based on minimizing the estimated filament thickness, along with Fourier space smoothing. 
We implement these techniques using two tools, \filgen\ which generates mock filaments and \filapt\ which analyses and processes them. 
We expect these tools to be useful in calibrating the performance of filament finders, thereby enabling searches for filament universality. 
}


\begin{document}
\maketitle
\flushbottom

\section{Introduction}
\label{sec:intro}
Matter in the Universe is arranged in an intricate and multi-scale pattern known as the ``cosmic web'' \cite{Bond_et_al1996}. This appears prominently in both observations (see, e.g. \cite{Lapparent_et_al1986, Colless_et_al2003}) as well as simulations (see, e.g., \cite{Jenkins1998, Illustris2014}). Despite being a highly non-linear structure, the skeleton of the cosmic web was laid down in the very early epochs of the Universe \cite{Zeldovich1970}. The regions of the cosmic web are broadly classified into voids, walls, filaments and nodes, and many properties of galaxies and dark matter halos are functions of their position in the cosmic web \cite{Aragon_calvo_et_al_2007a, Hahn_et_al2007b, Dubois_et_al2014, Kraljic_et_al2018}. Despite this, not all aspects of its structure and evolution are well-understood. In this work, we will focus on cosmic filaments, which are visually the most striking aspects of the cosmic web.

Cosmic filaments have been widely studied in the literature from multiple points of view (see, e.g., \cite{Colberg_et_al2005,Aragon-calvo_et_al2010a, NEXUS2014,Martizzi_et_al2019, Bonjean_et_al2018, Malavasi_et_al2020}). Their impact on the formation and evolution of low-mass halos is now well-established \cite{hahn+09,zomg-I,Paranjape_et_al2018a,musso+18}, and they likely form cosmic `highways' for the infall of gas onto halos, thus possibly aiding star formation \cite{Keres_et_al2005,Raychaudhury&Porter2005, Kirk_et_al2013, Konyves_et_al2015, Seth&Raychaudhury2020}. Despite decades of work, however, fundamental properties regarding the evolution and demographics (e.g., length and thickness distributions as a function of node halo properties) are not widely agreed upon \cite{Aragon-calvo_et_al2010a, 2pop2020,Zhu_et_al2021, wang+24}. Considering the well-known universality of density structure of dark matter halos \cite{NFW97, Einasto1965, merritt+06} and similar properties recently emerging for cosmic voids \cite{Pan_et_al2012, Hamaus_et_al2014, Nadathur_et_al2015}, it seems natural to expect the structure of cosmic filaments to also show a similar universality \cite{Yang_et_al2022}. The nature of this universality may have interesting implications for cosmological inference, especially for fundamental questions such as the nature of dark matter; e.g., the purported universal density profile for filaments may be sensitive to whether dark matter is cold or warm. Addressing the question of whether or not the filamentary structure is universal, however, is hampered by the complexity and multi-scale nature of the filamentary network, and a corresponding lack of agreement between the results of different approaches to studying it \cite{Libeskind_et_al2018}.

In this paper, we initiate a programme to systematically investigate and uncover possible universal features in the phase space structure of cosmic filaments. We will focus on filaments identified in the dark matter distribution, working primarily in real space. Although filaments identified in real data would use galaxies -- which are biased relative to dark matter and live in redshift space -- as tracers, our goal of uncovering universality is best approached with dark matter in real space (c.f., the universal real-space dark matter profiles of halos). The first step in such a study must be a robust characterization of the `spine' of a filament, akin to robustly identifying the `centre' of a halo. Errors in spine identification can potentially propagate into and bias the extraction of filamentary phase space profiles, and must therefore be understood for whichever filament finder one chooses to work with. Multiple filament finders exist in the literature, based on a wide range of approaches. For example, NEXUS \cite{NEXUS2013} is a grid-based method that identifies and classifies structures based on the Hessian of density or velocity shear fields, or the tidal tensor. \disp\ \cite{Disperse_theory, Disperse_illustration} is another widely used filament finder that is based on Morse theory. A detailed summary and comparison between filament finders is provided by \cite{Libeskind_et_al2018}.  Some filament finders such as \disp, T-ReX \citep{TRex2020}, COWS \citep{COWS2022}, Bisous \citep{Tempel_et_al2014}, etc directly provide the filamentary spines, whereas others such as NEXUS identify filaments to be 3D structures, which can be post-processed to obtain the spines. In this paper, we refer to the combination of all the analyses performed on the data to obtain the filamentary spines as a ``filament finder''.

Our point of departure is the realization that different filament finders identify rather different filaments of the cosmic web for the \emph{same} input data \cite{Libeskind_et_al2018}, and since the `true' spines are not known \emph{a priori}, there is no universally agreeable and principled method to select the `best' filament finder or calibrate its parameters (although, see \cite{Galarraga_Espinosa_et_al2023a} for one approach to calibrate \disp). If one already knew the `true' spines, this problem could be solved. Unlike halos and voids which can be described using spherical \cite{gg72} or ellipsoidal evolution \cite{bm96,smt01}, robust analytical models for the evolution of a filament, especially the shape of its spine, are harder to come by (although see \cite{stodolkiewicz63,ostriker64,ramsoy+21} for some analytical work relevant for gaseous filaments).

We circumvent this problem by the simple expedient of \emph{generating} a particle-based phase space realization (i.e., positions and velocities) corresponding to a \emph{specified} choice of spine shape and phase space profiles. The known choice of spine and profile shapes then serves as the `truth', where the filament spines will appear as perfect ridges in the density field \footnote{ In principle, one could imagine that our code could be modified so that the spine will correspond to a ridge in other physical fields of interest like the gravitational potential or the velocity divergence. However, this is beyond the scope of the current code.} , and the particle realization can be analysed using the filament finder of choice. We implement this filament generation using a tool \filgen\ that we describe later.

Even before applying any particular filament finder to such a particle realization, however, it is interesting to analyse potential causes for biases in recovered density and velocity profiles for \emph{perfectly known} spines. We identify one important source of bias, which occurs when the \emph{local radius of curvature of the spine} becomes comparable to or smaller than the filament thickness. We thus propose that a restriction to regions of low curvature (corresponding to relatively long, straight segments) is essential in obtaining unbiased results, \emph{regardless} of choice of filament finder. Next, we must consider the impact of noise induced by the filament finder in the recovery of the spine shape. We consider the publicly available \disp\ filament finder as a case study and evaluate the impact of noise on a variety of filaments generated using \filgen.\footnote{There is nothing special about \disp\ in this regard, except its ready availability and ease of use. In principle, we can repeat our analysis for any filament finder.} A standard approach to mitigate the biases in filament statistics induced by this `spine noise' (e.g., lengths are typically overestimated and density profiles are smoothed out) is to smooth the estimated spine curve using some method. We demonstrate that not only is smoothing essential, but the smoothing operation must be tuned to each individual filament, since the noise induced by the filament finder is, in general, a function of the intrinsic filament properties. To this end, we propose a novel optimization technique that is straightforward to implement and, as we demonstrate, successfully eliminates the biases induced by spine noise. Along the way, we also introduce a Fourier space smoothing operation with a low-pass filter, which we argue has some advantages over the commonly used smoothing over nearest neighbours in configuration space. We package all of these post-processing operations into a second tool \filapt\ which we also describe later.

The structure of the paper is as follows. Section~\ref{sec:tools} introduces the tools \filgen\ and \filapt, explaining in detail the assumptions made and operations performed. It also describes smoothing optimization and Fourier space smoothing methods. Section~\ref{sec:results} illustrates the performance of the tools at various positions in the input parameter space, and tests their robustness. The need for smoothing of filament spines, and the advantage of splitting by curvature are motivated. Section~\ref{sec:Applications} illustrates the advantage of using optimized smoothing, and also shows a simple case of studying filaments in redshift space. Section~\ref{sec:conclusion} concludes with a summary and discussion of possible improvements in and potential applications of \filgen\ and \filapt.  

\section{Tools and techniques}
\label{sec:tools}
Here, we present a set of two tools that can be used along with a filament finder to study different statistical properties of filaments in the cosmic web. The first tool -- the {\bf Fil}ament {\bf Gen}erator, or \filgen\ -- presents a way to generate the complete phase space information of particles in realizations of simple mock filaments, which can be used to test and calibrate filament finders. The second tool -- the {\bf Fil}ament {\bf A}nalysis and {\bf P}rocessing {\bf T}ool, or \filapt\ -- provides a way to analyse the filaments and extract their various profiles, provided their spines are well identified. We also introduce a new, parameter-independent, Fourier-based technique to post-process the spines identified by a filament finder, so that the profile estimation can be robust and unbiased. 

\noindent
\underline{\emph{Notation:}}
Throughout the paper, $r$ refers to the perpendicular distance from the $z$-axis in the cylindrical coordinate system. $z, \phi$ are the axial coordinate and the azimuthal angle in the cylindrical coordinate system respectively. The spherical polar coordinates (radius, polar angle, azimuthal angle) will be denoted by $(r_{sp}, \theta, \phi_{sp})$, and the Cartesian coordinates by $(x, y, z)$. The axis of the cylindrical coordinate system will always be aligned with the $z$-axis of the Cartesian coordinate system. The cubical box that is simulated has a side length $L$, and all the particles are assumed to have the same mass ($m_{\rm p}$). The expression $x_{N} \sim f(\alpha_i)$ means that $N$ independent values of the coordinate $x$ are drawn randomly from the distribution $f$, which is a function of the parameters $\alpha_i$. When there is no subscript $N$, it means that a single value is drawn. A uniform distribution between $[a,b]$ will be represented by $\mathcal{U}(a,b)$, and a Gaussian with mean $\mu$ and standard deviation $\sigma$ will be represented by $\mathcal{N}(\mu, \sigma)$.
The information generated by the filament finder as an output and the one accepted by the profile estimator as its input is in the Cartesian coordinate system. The units of length and velocities are \Mpch\ and \kms, respectively, although throughout the discussion, the units are irrelevant.

\subsection{Filament Generator \filgen}
\label{sec:filgen}
As mentioned in section~\ref{sec:intro}, multiple filament finders exist in the literature, each based on a different approach. These different filament finders identify somewhat different spines of the cosmic web. Since the ``correct'' spine is not known a priori, there is no robust way to select a particular filament finder or its parameters.  Hence, any statistic of the filament population that is evaluated is inherently linked to the choice of the filament finder and its parameters. This problem would be solved if one already knew the correct spine. Then one (or more) cost function(s) could be defined, and the filament finder and its parameters that minimize this cost would be preferred. Here, we present a filament generator \filgen, that produces realizations of simple mock filaments, that can be used for this purpose. Note that the multi-scale nature of the cosmic web can cause more complications in spine identification, and that is not accounted for in this case. However, in principle, one can generate a hierarchy of smaller and smaller filaments inside the main filament to account for the multi-scale nature. Aside from calibrating filament finders, this tool can be used to generate filaments with varied spines, density and velocity profiles. Then one can study the dependence of the biases in the inferred statistical properties on these parameters, and choose to select only those filaments in the simulations that give unbiased results. For example, we show how splitting the filaments by curvature and retaining only the low-curvature regions can give more robust and unbiased results. Also, since the velocity profiles are completely under the user's control, one can study the effect of the flows on the redshift space distorted filaments, as we shall illustrate using a simple toy example in section~\ref{sec:results}. 

The tool presented here generates a straight filament with a Gaussian cylindrical radial density profile, then curves it according to the given arbitrarily curved spine. One can specify the radial and longitudinal velocity profiles of the particles in the filament as well. One can add Navarro–Frenk–White (NFW) halos \cite{NFW97} at the ends of the filament, and model the halo outskirts based on the input density and velocity profiles in these regions. A uniform background can also be added if required. Finally, the complete phase space information of all the particles in the box is generated. Although the generation and study of the filament alone is enough in real space, modelling of the halos and their outskirts does play an important role when studying the filaments in redshift space. Also, the background and node modelling is important to create a realistic filament, to which a filament finder can be successfully applied to recover its spine.

We now discuss in detail the steps followed to generate the particles in various regions of the box, namely the filament, the nodes at its ends, the node outskirts and a uniform background.

\begin{figure}
    \centering
    \includegraphics[width=\textwidth]{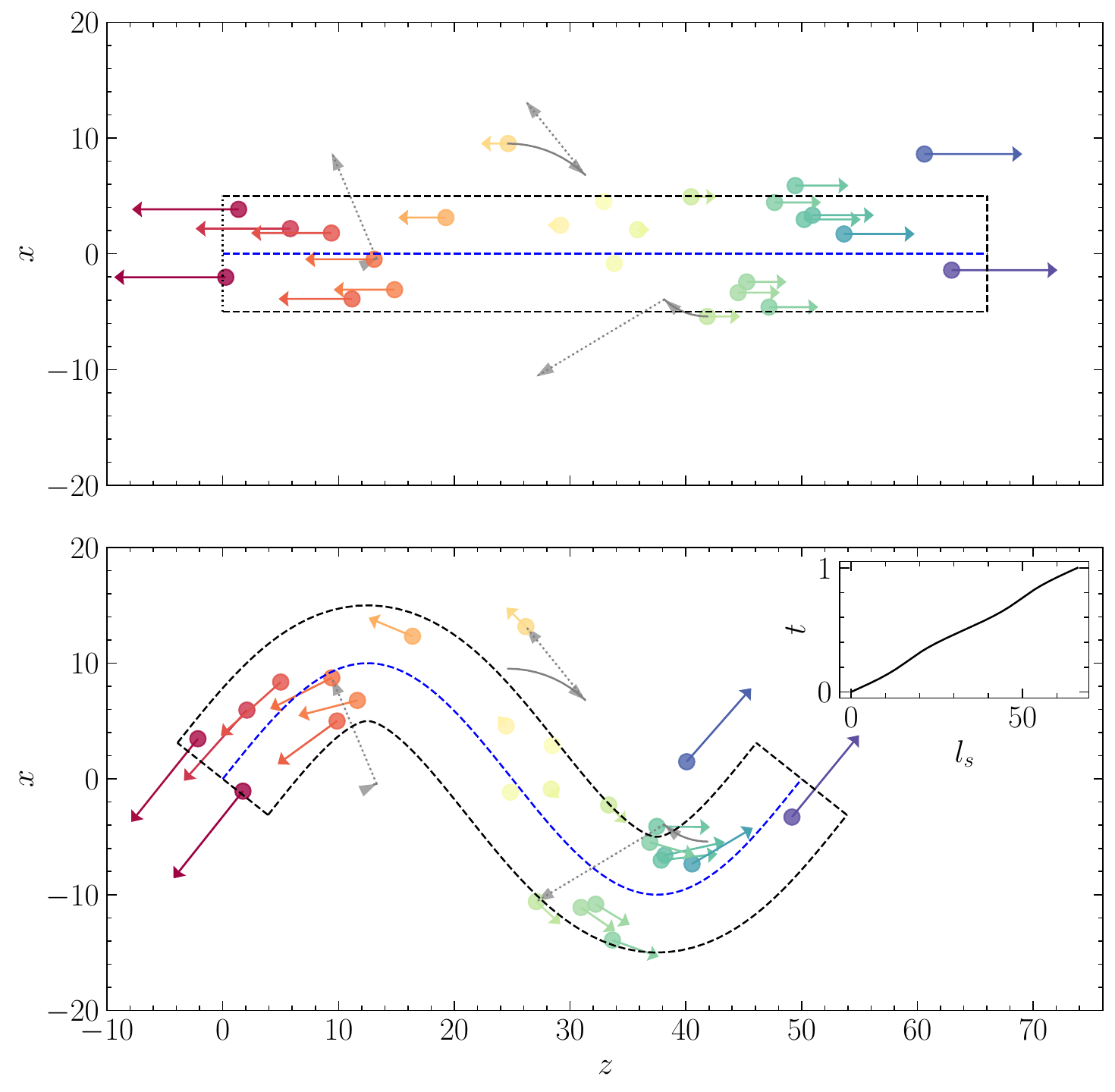}
    \caption{An illustration to demonstrate the generation of a filament. The two panels show the $y$-projection of the filament, and the filament spine in both cases has its $y$-coordinate to be identically zero. Top (bottom) panel shows the straight (curved) filament. The dashed blue line shows the spine of the filament, and the dashed black lines show the region at a radial distance of $r_f$ away from the spine. The markers show the position of the particles generated from a cylindrical Gaussian density profile. They are coloured by their $z$-coordinates before curving, so as to match the particles across the panels. The arrows show the velocity of the particles, which is assumed to have only longitudinal component for easy interpretation. The rotation and translation that curve the filament are shown for three of the particles using solid and dotted grey lines, respectively, the arrows indicating their direction. The bottom panel illustrates that the transformations are valid, retaining the radial and longitudinal profiles. The mapping between the length along the spine $l_s$ and the parameter $t$ used to define the spine of this filament is shown in the top right corner of the bottom panel. This mapping is necessary while curving a filament in \filgen. }
    \label{fig:cartoon}
\end{figure}
To model the filament, we assume that the filament density profile depends only on the perpendicular distance from the spine ($\rho(\vec{r})=\rho(r)$). The radial profile is assumed to be a truncated Gaussian:
\begin{align}
\rho(r) &= c_f\rho_0\exp{\left(-\frac{r^2}{2r_f^2}\right)} \,;\quad r\leq R_f\,,
\label{eq:Gaussian}\\
R_f & \equiv \sqrt{2r_f^2 \ln{(c_f)}} \,,
\label{eq:Gaussian_Rfdef}
\end{align}
where $r_f$ is the radius of the filament, and $c_f$ is the filament concentration defined to be the ratio of the density at the centre of the filament to the mean background density ($\rho_0$). The Gaussian profile is truncated at $R_f$, the radius at which the Gaussian density equals the background density. The cumulative distribution function (CDF) corresponding to this density profile is:
\begin{equation}
\label{eq: Gaussian_cdf}
{\rm CDF}(r) = \frac{ 1-\e{ -r^2/(2r_f^2}) }{1-\e{ -R_f^2/(2r_f^2)}}\,;\quad r \leq R_f\,.
\end{equation}
The velocity components $(v_r, v_z, v_\phi)$ are assumed to have Gaussian profiles, whose mean and standard deviation have to be specified by the user as functions of $\vec{r}$. One needs to describe the spine of the filament by a continuously differentiable parametric curve $\vec{P}(t)$ parameterized by $t \in [0,1]$. The tangent to this curve is given by $\vec{T}(t) \equiv d\vec{P}(t)/dl_s =( d\vec{P}/dt) (dt/dl_s)$, where $l_s$ is the length along the curve.

The process of creating the filament can be split into two broad tasks: (a) creating a straight filament with the given radial and longitudinal profiles, and (b) curving the filament according to the given spine; both of which we shall describe briefly. 
\begin{itemize}
    \item \textbf{Creating a straight filament}
    \begin{enumerate}
        \item Based on the input mean background density $\rho_0$, mass of the particles $m_{\rm p}$, and filament parameters $r_f$, $c_f$,  the number of particles $N_f$ making up the filament are calculated. From the given spine, the length of the filament $l_f$ is also calculated.
        \item The azimuthal and axial coordinates are generated as $\phi_{N_f} \sim \mathcal{U}(0, 2\pi)$ and $z_{N_f} \sim \mathcal{U}(0, l_f)$.
        \item To generate the coordinate $r$, uniform random numbers are generated between $[0,1]$, and these CDF values are converted to the corresponding $r$ values by inverting the CDF given in \eqn{eq: Gaussian_cdf}. 
        \item In case the velocity information is required, the means ($\mu_{i} (\vec{r})$) and standard deviations ($\sigma_{i}(\vec{r})$) of the Gaussians describing the velocity components at the position ($\vec{r}$) of each particle are evaluated from the input functions, where $i \in \{r, \phi, z\}$. 
        Then the velocity components are generated as $v_{i}(\vec{r}) \sim \mathcal{N}(\mu_i(\vec{r}), \sigma_i(\vec{r}))$. The set of cylindrical polar basis vectors form a local orthonormal basis; thus the velocity components are drawn from independent Gaussians with different dispersions.
        \item Appropriate coordinate transformations are used to get the position and velocity components in Cartesian coordinate system. This completes the generation of a straight filament.
    \end{enumerate}
    \item \textbf{Curving the filament}
    \begin{enumerate}
        \item A one-to-one mapping between the length along the spine ($l_s$), and the parameter $t$ is generated. Since initially the filament is aligned with the $z$-axis, the $z$-coordinate of the particle is the same as $l_s$. Thus, for any given $z_1$, there exists a unique $t_1$ and position on the spine ($\vec{P}(t_1)$) associated with it. Note that this mapping between the straight and curved spine preserves the length of the spine by construction.
        \item To curve the filament, two operations are performed on the position vector of every particle. First, a rotation $\mathcal{R}(t_1)$  centered on $[0,0,z_1]$ that takes the unit vector $\hat{z}$ to $\hat{T}(t_1) \equiv \vec{T}(t_1)/\lVert \vec{T} (t_1) \rVert$, which is the unit vector along the required tangent at $z_1$. Second, a translation $\mathcal{T}(t_1)$ that takes the point $[0,0,z_1]$ to $\vec{P}(t_1)$. 
        Note that the exact same operations need to be performed on all the particles lying in a plane perpendicular to the spine at any given value of $z$, and hence $\mathcal{R}, \mathcal{T}$ are a function of the $z$-coordinate of the particle alone (and not $x, y$). The velocity vectors only need to be rotated, and not translated.
    \end{enumerate}
\end{itemize}
Finally, the curved filament is translated to the specified starting point inside the box. Figure~\ref{fig:cartoon} illustrates this procedure for a toy filament. The rotation and translation is shown for three random points using grey curves. The mapping between $l_s$ and $t$ for this filament is shown in the top right corner of the bottom panel. Note that although we have restricted ourselves to a Gaussian radial density profile here, it is trivial to change this to an arbitrary radial profile, as long as the CDF is well-defined. Imagine the initial straight filament to be made up of thin discs placed adjacent to each other, which are to be
rotated in accordance with the local direction of the final curved spine in order to obtain the curved filament. When the curvature is low, the discs do not affect each others’ profiles substantially after rotation, thus retaining the radial and longitudinal profiles. This is exactly what is being done in our case.

We assume that the nodes at the ends of the filament are NFW halos truncated at $R_{200c}$, where $R_{200c}$ is the radius at which the mean enclosed density is $200$ times the critical density $\rho_{\rm{crit}}$ of the Universe. 
The density profile is given by:
\begin{align}
\rho_{\rm{NFW}}(\rsp) \propto & \, r_{\rm sp}^{-1}\,(r_{\rm sp}+r_{\rm s})^{-2}\,,
\label{eq:rho_NFW}
\end{align}
where $r_{\rm s}$ is the scale radius (with the usual concentration parameter being $c\equiv R_{\rm 200c}/r_{\rm s}$), and the normalization is chosen so as to enclose a density of $200\rho_{\rm crit}$ inside the radius $R_{\rm 200c}$. 
To generate the position coordinates of the particles of a given halo, the number of particles $N_h$ are calculated from the given $m_{\rm p}$, $R_{200c}$ and $\Omega_m$. Then, the angular coordinates are generated as $N_h$ realizations of $\phi_{\rm{sp}} \sim \mathcal{U}(0, 2\pi)$ and $\mu   \sim \mathcal{U}(-1,1)$, 
where $\mu \equiv \cos(\theta)$. The radial coordinates \rsp\ are generated using inversion on the tabulated CDF of the NFW profile.

To get the velocity of the particles, we follow the approach described in the appendix of  \cite{Shethetal2001}, modifying the equations to include a constant velocity anisotropy $\beta$ \cite{Binney&Tremaine1987}, where we solve the Jeans equation for an NFW halo:
\begin{equation}
    \frac{\partial
    (\rho_{\rm{NFW}}\sigma_{\rsp}^2 )}{\partial \rsp}+ \frac{2 \rho_{\rm{NFW}}}{\rsp}\beta \sigma_{\rsp}^2+\rho_{\rm{NFW}}\frac{G M(< \rsp)}{\rsp^2}=0 \,,
\end{equation}
to obtain the radial and tangential velocity dispersion $\sigma_{\rsp}, \sigma_t$ respectively.  Here $M(<\rsp)$ is the mass enclosed within the radius $\rsp$. We assume $\sigma_\theta^2=\sigma_{\phi_{\rm{sp}}}^2 = \sigma_t^2/2$, and $\sigma_t^2 = 2\sigma_{\rsp}^2(1-\beta)$. Here, $\beta$ quantifies the velocity anisotropy of the halo, and can take values between $(-\infty, 1]$, with $\beta=0$ representing isotropic velocity dispersion. We assume that $\beta$ is independent of $\vec{r}$, and one needs to specify its value as an input. In principle, this can be extended to include realistic anisotropy profiles that might be relevant for alignment statistics of halos with filaments (see, e.g., \cite{Catelan_et_al1996, Jeeson-Daniel_et_al2011, Ramakrishnan_et_al2019}), but we do not pursue this here. Similarly, we leave the modelling of halo shapes and their tidal alignment, which are potentially relevant for weak-lensing studies (see, e.g., \cite{Blazek_et_al2011, Maion_et_al2023}), to future work. The differential equation is solved numerically with the boundary condition $\sigma_{\rsp}^2=0$ at $\rsp= R_{200c}$, to obtain $\sigma_{\rsp}$ as a function of $\rsp$. The velocity components for each particle with position $\vec{r}$ is generated as $v_i \sim \mathcal{N}(0, \sigma_i)$, where $i \in \{\rsp, \theta, \phi_{\rm{sp}} \}$. The position and velocity coordinates are transformed into Cartesian coordinates, and the halos are translated so that their centres coincide with the end points of the filament.

We assume that the outskirts of a given halo are defined between $R_{200c} < \rsp \leq 4R_{200m}$,\footnote{$\sim 4R_{200m}$  defines the tidal environment of a halo, as motivated by \cite{Paranjape_et_al2018a, Ramakrishnan_et_al2019}. } where $R_{200m}$ is the radius at which the mean enclosed density is $200$ times the mean density $\rho_0$. The generation of density and velocity coordinates of the particles in this region is in principle similar to that of the generation of the NFW halos. The CDFs corresponding to the density profiles in these regions have to be specified as an input, along with the velocity mean and dispersion profiles as functions of $\vec{r}$. One also needs to specify the mean density in this region. First, the number of particles in the halo outskirts is calculated from the mean density. The position coordinates are generated following the steps used in NFW halos, except the CDF is different. The velocity profiles are assumed to be Gaussians with given means and standard deviations. The velocity components are generated accordingly. The rest of the steps are the same as in the case of the NFW halos. After the creation of the halos and their outskirts, the parts of the filament lying inside the halos are deleted, so that the filament ends at the edges of the halos; however, we retain the filament sections lying inside the halo outskirts. The choices for the density and velocity profiles of halo outskirts in the present work are arbitrary, since accurate characterizations of these, while correctly accounting for mid- to large-scale density and velocity correlations, do not yet exist (although see \cite{Tinker_et_al2005, Bosch_et_al2013}; in halo model language, this is the so-called 1-halo to 2-halo transition regime). In future work, we plan to characterise these regions using halo environments measured in simulations in order to improve this part of \filgen. 

Finally, a background is added. We assume that the background particles are distributed uniformly with a density $\rho_0$, and have isotropic velocities. The position and velocity Cartesian coordinates of the particles are generated as $x_{N_b}, y_{N_b}, z_{N_b} \sim \mathcal{U}(0, L)$ and $v_{x,N_b}, v_{y,N_b}, v_{z, N_b} \sim \mathcal{N}(0, \sigma_b)$, where $N_b= \rho_0 L^3/m_p$, and $\sigma_b$ is the $1$-dimensional isotropic background velocity dispersion. All the background particles lying within a distance $R$ from the centres of the two halos are deleted. Here, $R=R_{200c}$ of the respective halos if the halo outskirts are not modelled, otherwise $R=4R_{200m}$ of the respective halos. All the background particles lying within a perpendicular distance less than the truncation radius $R_f$ (see equation~\ref{eq:Gaussian_Rfdef})  from the filament spine are also deleted. To calculate the distance to the spine, the spine is very finely sampled, and the distance for each background particle to all these sample points is calculated. The minimum of these distances is taken to be the distance of the particle to the spine. The deletion of the particles is to ensure that the filament, halos and their outskirts have the expected profiles (e.g. NFW in the case of halos), and no extra constant component.

\begin{figure}
    \centering
    \includegraphics[width=\textwidth]{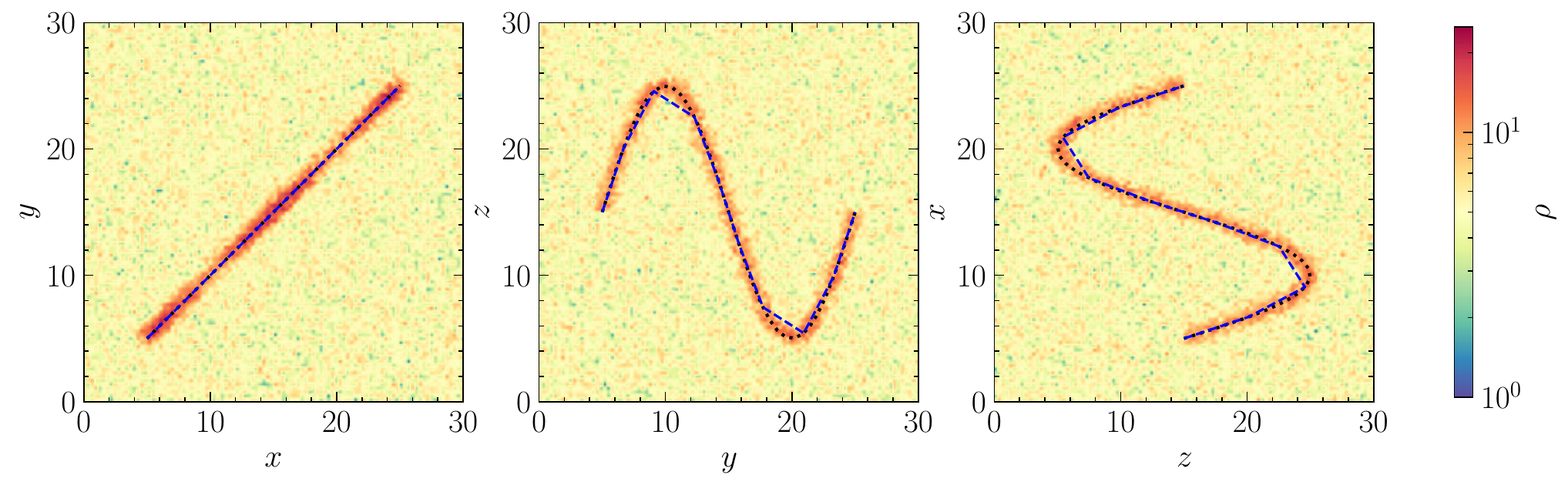}
    \caption{Plots showing number density projections of the fiducial filament along the three Cartesian axes for one realization. All axes are in units of \Mpch, and the number density is given in units $(h/\rm{Mpc})^3$, with input background density being $5$. The radial density variation in the filament is not visually apparent because the radius of the filament is small. The apparent density variation along the length of the filament in the left panel is a purely projection effect. Black dotted curves show the actual spine of the filament. Blue dashed curves show the spine when traced using sparse tracers (see section~\ref{sec:spine_sampling} for details). We see that in the case of sparse sampling, the spine is not robustly traced in regions of high curvature.}
    \label{fig:filament_illustration}
\end{figure}

Now that all the regions have been modelled, the position and velocity information of all the particles in the box is combined and returned at the output. For illustrations, see Figure~\ref{fig:filament_illustration}, which shows the density projections along the three Cartesian axes of a filament generated using this tool. Here, we have set the node masses and radii to zero (i.e., no node halos), and modelled the halo outskirts to be identical to the uniform background (i.e., no halo outskirts). 

\subsection{Filament Analysis and Processing Tool \filapt}
\label{sec:filapt}
Here, we present a tool that can be used to estimate various statistical properties of a filament as a function of the perpendicular distance from its spine and length along the spine. This tool takes the phase space information of all the particles of interest, and a discretely sampled spine as its inputs, and generates density and velocity (mean and dispersion) profiles. This tool can also split the filament based on the curvature of the spine $\kappa$ (see Appendix~\ref{app:curvature} for details), and statistics of interest can be computed separately for regions of different curvature. This can substantially clean up the profiles, and will be useful in getting better statistical information, as explained in section~\ref{sec:results}.

Since many widely used filament finders such as \disp\ identify a discretely sampled spine, \filapt\ provides convenient methods to estimate some of the most basic and important properties of filaments immediately after applying the filament finder. We shall apply this tool to various filaments generated by using \filgen\ to study the applicability and robustness of both the tools in section~\ref{sec:results}.

Here, we assume that the filament is made up of tiny straight cylinders whose axes are the line segments joining consecutive points of the discretely sampled spine. This is a good approximation as long as the length of the line segment at a point is substantially smaller than the radius of curvature $R_\kappa=\kappa^{-1}$ of the spine at that point. We shall elaborate on this point in section~\ref{sec:results}. In case the number of particles in the box is high, we provide a way to sort them into smaller sub-boxes; the code then processes only a subset of particles that lie in the sub-boxes close to the filament spine, resulting in a substantial speed-up. Once the phase space information of all the particles in the box along with the filament spine are provided to \filapt, the following operations are performed by the module \filapt\texttt{.ExtractProfiles}:
\begin{enumerate}
    \item The code starts by defining radial bins (logarithmically spaced by default), based on the user-defined parameters. The minimum and maximum radii of interest are $r_{\rm{min}}, r_{\rm{max}}$ respectively. Then, it goes over each line segment making up the spine one at a time, and performs the same operations as follows accounting for the boundary conditions when necessary.
    \item A rotation is performed on both the positions and velocities of all the particles, which takes the vector directed along the segment to the $z$-direction. Then, the coordinate system is translated so that the starting point of the segment lies at the origin. In the case of periodic boundary conditions, the particles are ``unfolded'' accordingly.
    \item All the particles with $r_{\rm{min}} \leq r \equiv \sqrt{x^2+y^2} < r_{\rm{max}}$ and $0\leq z < d_s$ (where $d_s$ is the length of the segment) are selected. These form a cylinder around the segment. The cylinder is translated so that its starting point coincides with the end point of the previous segment.
    \item When all the segments are processed, one is left with a straightened filament in the form of a cylinder, with its axis lying along the $z$-direction. This is chopped into 3D bins of $(r, z, \phi)$. The density $\rho$, mean radial and longitudinal velocities $\langle v_r \rangle, \langle v_z\rangle$, and their corresponding dispersions $\sigma_{r}, \sigma_{z}$ are calculated for each bin. This gives the 3D profiles for the filament. Note that the number of longitudinal bins is independent of the initial sampling of the spine, as the bins are created after straightening the filament. 
    \item In order to calculate the 1D radial (longitudinal) profiles, all the particles in the given $r$ ($z$)-bin are considered irrespective of their $\phi$ and $z$ ($r$) coordinates. The statistics of interest are evaluated from the data of all the particles in that bin.
    \item When splitting by curvature is required, the $\kappa$ values for all the segments are calculated at the beginning using equations~\eqref{eq:kappa_f}-\eqref{eq:kappa_final} by the module \filapt\texttt{.ComputeCurvature}. The segments are split into appropriate groups selected by $\kappa$ and the 1D radial profiles are calculated for each of the groups separately. This is useful for studying the effect of curvature on the radial profiles, as we discuss in section~\ref{sec:effect of radius}. 
\end{enumerate}
It should be noted that this method does not conserve mass when the filament spine is curved, in that a given particle may be counted more than once when estimating a profile. This is because, in the case of a curved spine, the same region slightly away from the spine can be associated with multiple points on the spine, and will be counted more than once in the process of unravelling. However, since we are interested in the field values as a function of distance perpendicular to and along the spine, this is the correct way to estimate the profiles.

One can estimate these profiles for multiple comparable filaments, and stack them to get more robust statistics. \filapt\ is fast, taking only a fraction of a second to calculate all the profiles of interest for a typical cosmological dark matter filament. As we shall illustrate in the next section, \filapt\ gives robust and unbiased estimates of the profiles, especially in regions of low $\kappa$, provided the spine determination and sampling is robust enough.

\subsubsection{Spine processor /smoother}
\label{sec:spine processor}
We have stressed in the discussion so far that a robust estimation of the spine is necessary for the extraction of unbiased profiles. However, it is a widely acknowledged fact in the literature that even in the most ideal of circumstances, the filament spines recovered by any filament finder will have errors. For particle-based filament finders, these errors originate from both the imperfections in the method of filament finding, as well as the discreteness of the tracers used. For grid-based filament finders, they originate from the finite size of the grid and Poisson noise in the simulations used to estimate the fields on the grid itself. Note that the errors will persist even when the parameters of the filament finder are optimally calibrated. These errors on the spine will in turn lead to systematic errors in the estimation of various profiles of the filaments (since the recovered spine is not exactly aligned with the actual spine, the inferred radial and tangential directions will not be exact), as well as in estimates of the filament lengths. Thus, it is essential that the spines identified by filament finders are processed properly before being used to calculate any statistics.

One of the most widely used methods of smoothing the spines obtained from particle-like data using \disp is to average the position of every point on the spine by those of its immediate neighbours, keeping the end points fixed \cite{Flows_around_galaxiesI, ramsoy+21}. This is done iteratively \Nsm\ number of times, until the filament appears smooth enough, or matches with the visually apparent filaments.  Thus, the number of iterations is highly subjective. Also, if the same value of \Nsm\ is used for all the filaments (as is common), the filaments with less number of points on the spine will invariably be over-smoothed as compared to the other filaments. Another problem with this kind of smoothing is that the outlier points with large errors tend to pull the other points progressively away from the actual spine with each iteration, when \Nsm\ is small. However, this effect vanishes at high enough \Nsm. On the other hand, there is no convergence with increasing \Nsm\ (i.e., more smoothing), since the smoothed spine at large enough \Nsm\  always tends to a straight line joining the end points, regardless of the actual original shape. 

Here, we present a new approach that optimizes the smoothing parameters for the chosen smoothing method separately for each individual filament. The idea is to iteratively smooth the filament by varying the smoothing parameters and estimating the density profile after each iteration. Both under- and over-smoothing will tend to broaden the density profile, and thus the narrowest density profile should correspond to the optimum smoothing. This removes any ambiguity related to the choice of smoothing parameters, as well as optimally smooths every filament separately. We also present an alternative method to the neighbour smoothing: Fourier smoothing, where the spines are processed in Fourier space by applying a low pass filter. We explain both the Fourier smoothing and optimization in Appendix~\ref{app:spine_smoothig}, and illustrate the improvement provided by optimization in section~\ref{sec:optimized_smoothing}. The smoothing of filament spines using optimized Fourier smoothing is implemented by the module \filapt\texttt{.SmoothSpine}. Although we illustrate our results using spines obtained via \disp\, and compare our Fourier smoothing with neighbour smoothing which is widely used by \disp\, users, the optimized smoothing introduced in this paper can be used on discrete spines obtained by any filament finder. We also note that there exist other techniques to smooth the filament spines such as the one employed in T-ReX, the robustness of which we have not explored in this paper.

\section{Results and Discussion}
\label{sec:results}
Here, we illustrate the results of \filgen\ and \filapt\ using some examples. We generate different filaments with densities comparable to those expected in a $600\, \Mpch$ N-body simulation with $1024^3$ particles. The velocity profiles are also similar to what one might expect in typical cosmological filaments (see section~\ref{sec:fiducial} for details). First, we illustrate a thin filament with a sinusoidal spine, in order to sample regions of varied curvature. The spine is finely sampled, with zero noise to study the working of the profile generator in ideal conditions. This is our fiducial model. Then, we generate a thick filament, where $r_f$ is equal to $R_\kappa$ of the spine at the point of its maximum curvature, and illustrate various interesting effects that appear in such situations. Next, we use the fiducial filament, but with a sparsely sampled spine and illustrate the advantages of curvature splitting. For all the aforementioned examples, the spine provided to the estimator was perfect, with no error and noise, which is far from what one expects in real situations. To account for realistic noise, we estimate the profiles of the fiducial filament by using a spine identified using \disp. We show that this leads to unsatisfactory estimations of the profiles, and there is a need to post-process the noisy spines. We do this using the two methods, and compare the results. 

\begin{figure}
    \centering
    \includegraphics[width=\textwidth]{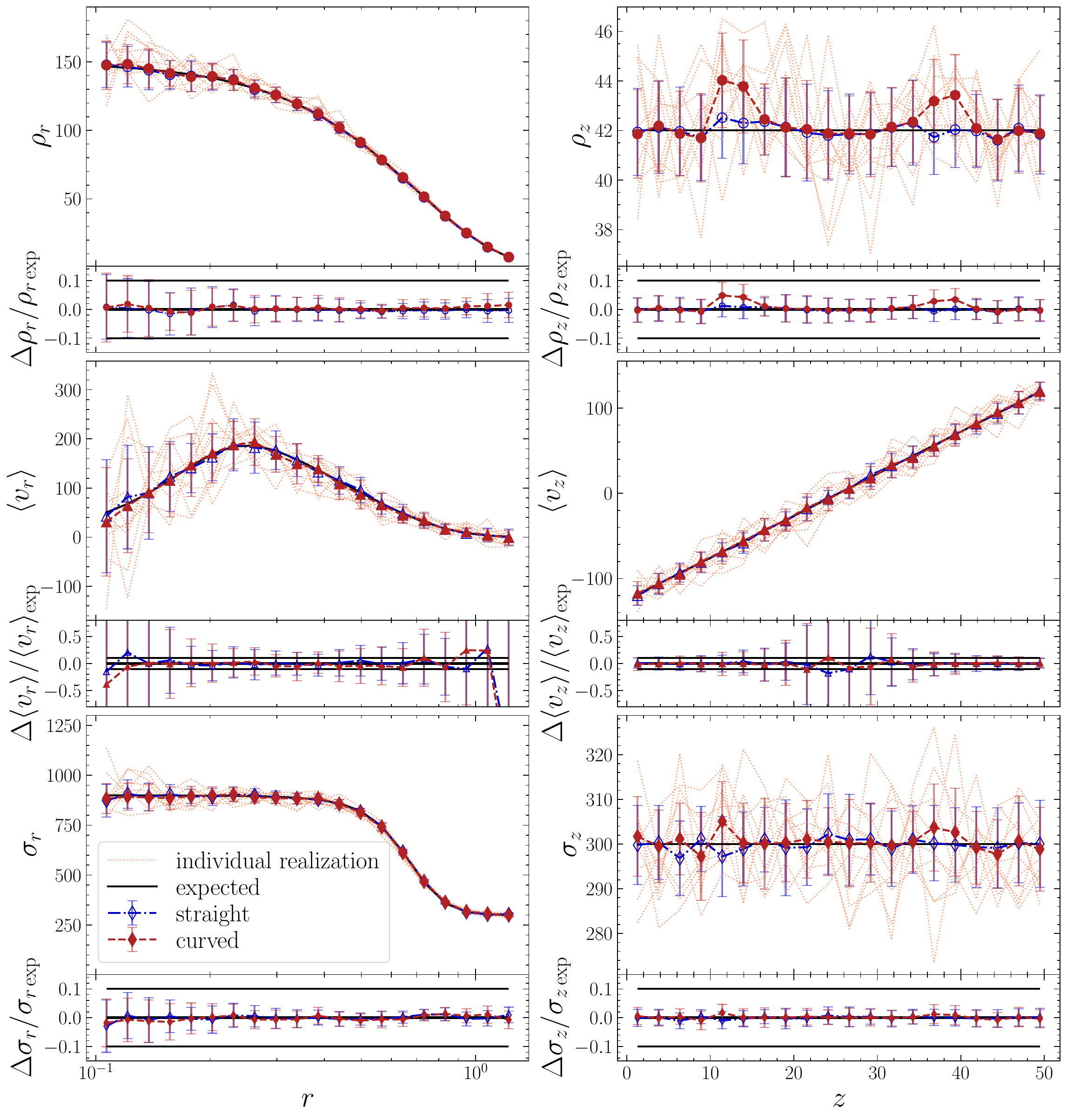}
    \caption{Various profiles for the fiducial filament model. The left (right) panels show longitudinally (radially) averaged radial (longitudinal) profiles. All horizontal axes are in units of \Mpch. See the text for a detailed explanation. The top, middle and bottom left (right) panels show the density, mean radial (longitudinal) velocity and radial (longitudinal) velocity dispersion profiles respectively. Densities have units $(\Mpch)^{-3}$, and velocities are given in \kms. The markers and error bars show the mean and standard deviations across $50$ realizations. Solid red markers represent the fiducial curved filament, whereas open blue markers represent the corresponding straight filament. 10 random realizations of the curved filament are shown with red dotted lines for reference. The solid black lines represent the expected analytical profiles. The narrow panels below each main panel show the relative errors with respect to the expected profiles to give better intuition about the error bars. Thin horizontal black lines in these panels show the $\pm 10 \%$ levels. We see that there is an excellent agreement between the generated and the analytical profiles.}
    \label{fig:fiducial}
\end{figure}

\subsection{The fiducial filament}
\label{sec:fiducial}
We generate an aperiodic box of side $30\,\Mpch$, with a curved filament having $r_f=0.5\,\Mpch$, and $c_f=30$. Since we are working in real space at this stage, there is no need to model the node halos and their outskirts, as they will
be cut off in any case while estimating the profiles. Therefore, for all the real space examples we only model the filament and the background. The region of maximum curvature of the spine has $R_\kappa = 2\, \Mpch$, so that $R_\kappa \gg r_f$. Hence, the effect due to the overlap of particles because of curving of the spine will be negligible, and we can safely compare the profiles of the curved filament with the corresponding inputs. Figure~\ref{fig:filament_illustration} shows the density projections of this fiducial filament along the three Cartesian axes. Note that the density variations seen along the spine in the left panel are purely projection effects.

Although considerable work has been done in literature to study the density profiles of cosmological filaments see, e.g., \cite{NEXUS2014, Colberg_et_al2005, 2pop2020, Yang_et_al2022, Galarraga_Espinosa_et_al2023a}), not much attention has been given to their velocity profiles (although, see, \cite{Rost_et_al2024, ramsoy+21}). However, it is expected that the matter flows radially onto the filament, and longitudinally away from its centre towards the nodes. Thus, we adopt the following ansatz for the velocity profiles: 
\begin{align}
    \langle v_z \rangle & = V_{0z} \times \frac{(z-l_f/2)}{l_f} \,, \label{eq:vz_fil}\\
    \langle v_r \rangle & = \frac{V_{0r}}{2}\bigg[1+ \erf{c\left(r/R_f-a\right)}\bigg]\,\bigg[1+\erf{b\left(a-r/R_f\right)}\bigg]\,, \label{eq:vr_fil}\\
    \sigma_r &= \sigma_0\big[2-\tanh(u(r/R_f-g))\big]\,.\label{eq:sigvr_fil}
\end{align}
Here, $l_f$ is the length of the filament, and $\langle v_i \rangle, \sigma_i$ are the mean and standard deviation profiles of the velocities, where $i \in \{ z, r\}$, and $\erf{x}$ is the error function. $\sigma_z$ is assumed to be constant along the spine, with a value $300 \,\kms$, same as that for the background 1D dispersion. Here, we assume $V_{0z}=V_{0r}=250 \, \kms, \sigma_0= 300\,\kms, a=0.125, \, b=2.5, \,c=15,\, u=8,\, g=0.5$, which approximately match the results of \cite{Rost_et_al2024} for the radial velocity profiles. The particular form of $\langle v_r \rangle$ is chosen to get a profile shape in which the mean infall peaks at some particular $r$ slightly away from the filament spine, and smoothly approaches $0$ at both the extremes of $r$, the decrease in $\langle v_r \rangle$ being shallower in the outer regions of the filament as compared to the inner ones.

We generate $50$ realizations of this filament, and estimate the density and velocity profiles for each one separately, taking a finely sampled spine with $50$ equi-distant points along its length. We have checked that our results for the fiducial filament are converged with respect to increasing these numbers. The mean and error profiles are calculated to be the mean and standard deviation across the realizations. The velocity dispersion statistics are calculated for $\sigma_i^2$, and error propagation is used to get the error on $\sigma_i$. The maximum radius probed is $R_f$, so that any longitudinal profile as a function of $z$ is averaged over a disc of radius $R_f$ perpendicular to the spine at the location $z$. The same procedure will be followed for all the other examples, and we shall only mention the quantity or parameter that is changed relative to the fiducial filament.

Figure~\ref{fig:fiducial} shows the 1D radial (left panels) and longitudinal profiles (right panels) for the fiducial filament, as well as the corresponding straight filament. The expected (input) profiles are shown as solid black curves. We have also shown $10$ randomly chosen individual realizations with dotted curves, to give an idea of the variation across realizations. We also show the relative errors on the profiles, for better visualization of the errors. As seen from the plots, the recovered mean profiles are in excellent agreement with the input profiles. However, note that the individual profiles can be noisy; stacking of comparable filaments would clearly be advantageous.  The only noticeable deviation of the mean profiles from the expected ones is the slight increase in the longitudinal density profiles at two points along the spine, which correspond to regions of highest curvature. We will return to this point below.

Even though the mean values are slightly higher, they are still in agreement with the input profile within the error bars. Another point worth mentioning is that the errors on both the density and velocity dispersion profiles are almost always well within $10 \%$. The large errors on $\langle v_z \rangle$ near the centre of the filament are not because of anything physical, but because the denominator approaches zero at this point.  Also, the large errors on $\langle v_r \rangle$ closer to the spine (small $r$) are because the radial velocity dispersion is high in this region. The increase in errors away from the spine is, again, because the denominator approaches zero. Thus, we have shown that both  \filgen\ and \filapt\ perform extremely well when $r_f$ is much smaller than the maximum $R_\kappa$ of the spine, and the filament spine is well estimated.

\subsection{Effect of filament radius}
\label{sec:effect of radius}
\begin{figure}
    \centering
    \includegraphics[width=\textwidth]{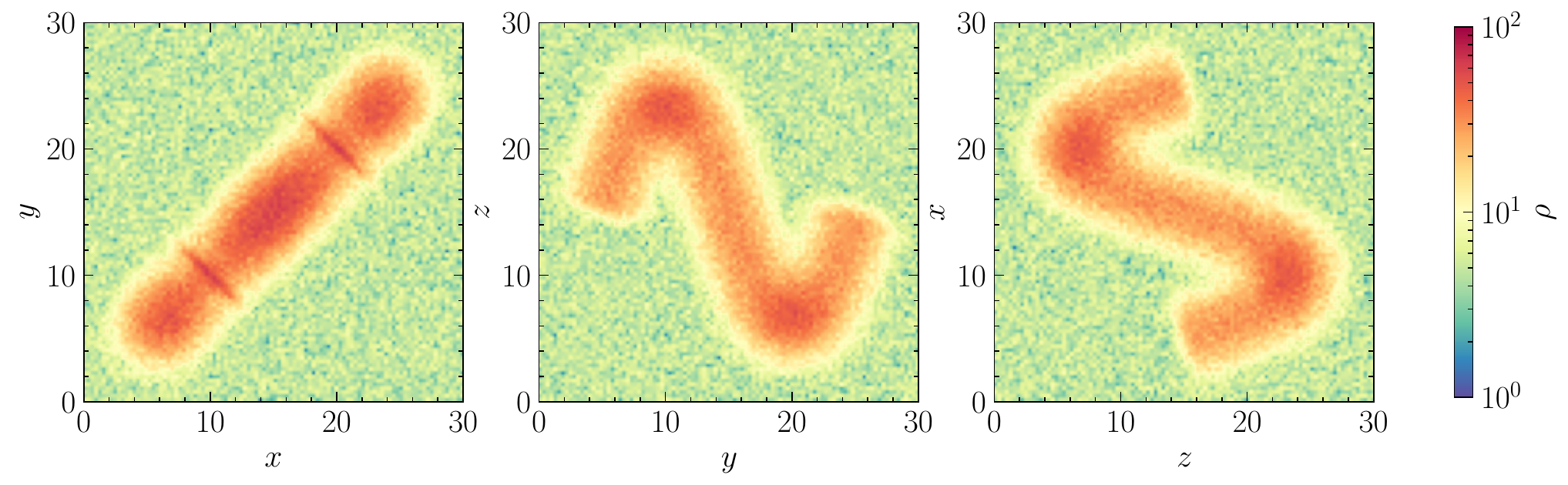}
    \caption{Projected number density for the thick filament. The panels are the same as Figure~\ref{fig:filament_illustration}. With larger $r_f$, the radial density variation in the filament is clearly visible. The density variation along the spine in the left panel is a projection effect, but that in the other two panels is real. See section~\ref{sec:effect of radius} for details.} 
    \label{fig:fat_fila_illustration}
\end{figure}

\begin{figure}
    \centering
    \includegraphics[width=\textwidth]{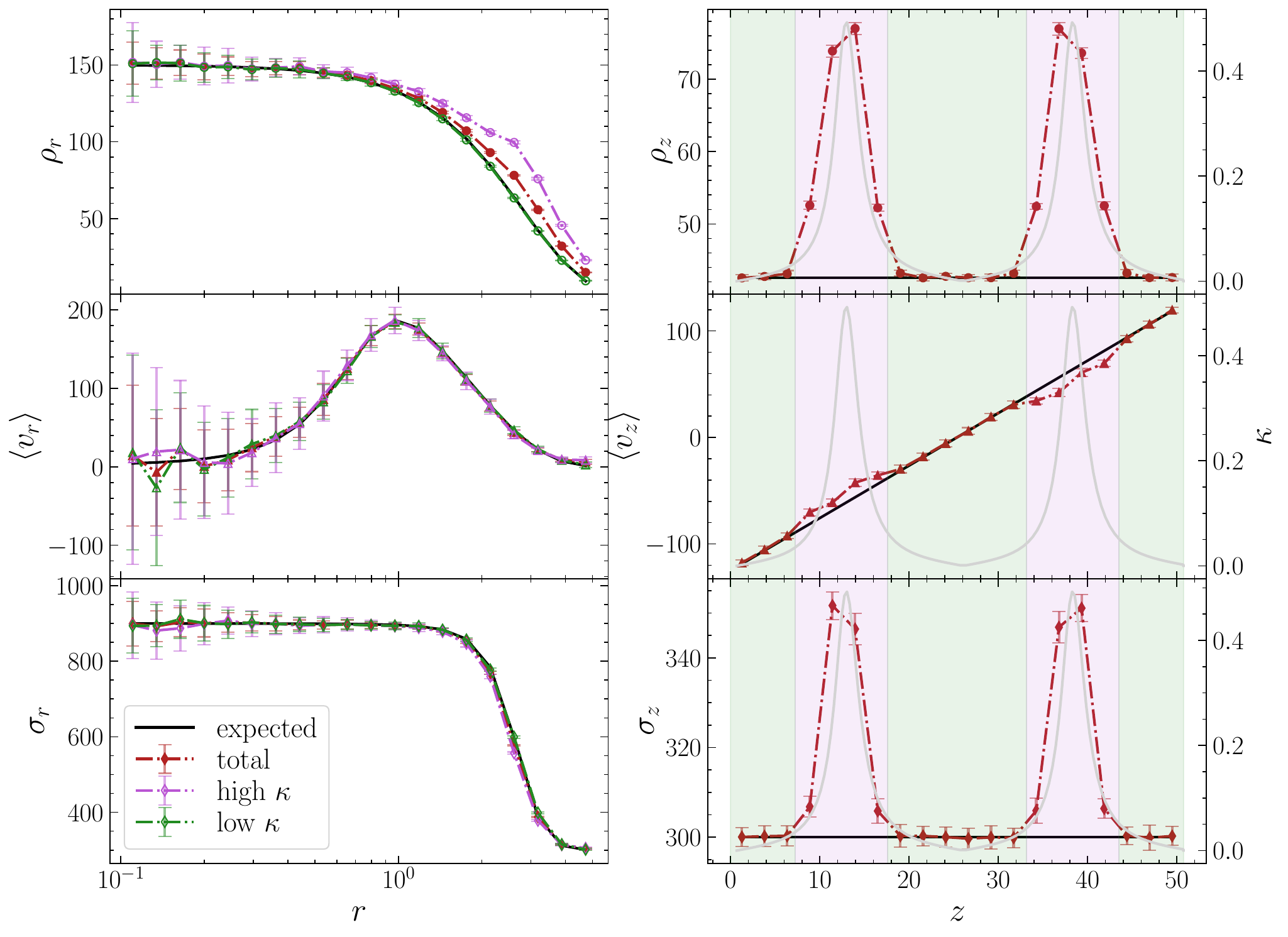}
    \caption{Radial and longitudinal density and velocity profiles for the thick filament. The panels are the same as in Figure~\ref{fig:fiducial}; except the error panels are skipped, as the error bars are qualitatively similar to the fiducial case. The curvature $\kappa$ of the filament is over-plotted in grey solid lines in the right panels. The regions of high (low) $\kappa$ are shaded with purple (green). Red colour represents the profiles for the whole filament. Purple (green) colours show the average radial profiles obtained for segments having high (low) $\kappa$. We see that the major contribution towards the biasing of the density profile arises from the high curvature regions.}
    \label{fig:curvature_split}
\end{figure}

As we saw in the case of the fiducial filament, the regions of high curvature show some effects worth studying. Since we first create a straight filament with the given profiles and then curve it, there will be an overlap in the regions where $R_\kappa \sim r_f$, creating deviations from the expected profiles. Therefore, one has to be cautious while using \filgen. Here, we illustrate these effects using a filament identical to the fiducial one, except now $r_f=2 \,\Mpch$, which is equal to the minimum $R_\kappa$ of the spine. Figure~\ref{fig:fat_fila_illustration} shows the projected density for this thick filament. It is evident that overlap in regions of high curvature is giving rise to deviations from the input density profile. We now study the profiles quantitatively in detail. 

Figure~\ref{fig:curvature_split} shows the density and velocity profiles of the thick filament. The panels are identical to those shown for the fiducial filament, except that we do not show the residuals since these are qualitatively similar to those obtained for the fiducial filament. The black curves show the expected profiles whereas the red markers show the profiles obtained from the realizations constructed using \filgen. In the right panels, the curvature $\kappa$ of the spine as a function of length along it is overlaid on the profile plots. It is seen that regions of high $\kappa$ (shaded purple) show substantial deviation from the expected profiles, whereas in regions of low $\kappa$ (shaded green), there is a good agreement between the expected and recovered profiles. Although the deviation is most prominent in the longitudinal profiles, one can also see corresponding deviations in the radial density profile. The effect is more prominent in the outer regions of the filament, where the overlap is more.

In order to recover unbiased profiles we split the spine into two sets comprising of segments of high ($\kappa \geq \kappa_{\rm th}$) and low curvature ($\kappa <\kappa_{\rm th}$), with the threshold set to $\kappa_{\rm th}=1/(10r_{f})$.
The high and low curvature regions are represented by purple and green shaded regions respectively in Figure~\ref{fig:curvature_split}. The radial profiles are calculated separately for both the sets, and are plotted in the left panels. As expected, the high curvature regions show maximum deviation from the expected profiles, and the low curvature regions recover the expected profiles perfectly. Also, note that the decrease in signal-to-noise ratio is not substantial (here, the volume of the low curvature regions is $\approx 60 \%$ of the total volume, leading to the signal-to-noise ratio of $\approx 0.77$ times the original). \emph{Therefore, removing the high curvature regions can substantially increase the accuracy of the profile estimate without significantly affecting the precision.} In general, rather than discarding the high curvature region, it will be beneficial to study the different curvature regions separately. 

We would like to emphasize that the effects discussed here are a result of the way the filaments were constructed in the first place: starting with a straight filament with the expected profiles, and then curving it. This will not always be so in the case of real cosmological filaments. The spines may be inherently curved and the radial growth with accretion might occur on this curved filament. We do not presume that in such cases the high-$\kappa$ regions will necessarily lead to higher densities, but differences in accretion rates and possibly other physical processes might be expected in such regions. However, Figure~$8$ in \cite{Galarraga_Espinosa_et_al2023a} shows that filament spines can, in fact, curve over time. In such cases, one might expect effects similar to what we have illustrated here. In any case, it appears prudent to split the spines by curvature to account for curving after filament generation, or different accretion due to different geometry. This will increase the robustness of the inferred filament profiles. Also, we show below that the high-$\kappa$ regions are most affected due to sparse sampling of the spine as well as noise; in these cases as well, splitting by curvature gives more unbiased and robust results.

Another point to note is that when $r_f \gtrsim R_\kappa$, although the filaments generated using \filgen\ do not exactly recover the input profiles, they are still perfectly valid filaments. While using these, however, the correct profiles will no longer be the input analytical profiles; instead, they have to be determined by using a very finely sampled spine, and numerically calculating the profiles using \filapt.

We would like to point out that the curvature threshold is set to $\chi/r_f$, where the choice of the factor $\chi$ is not fixed based on any physical argument. We have checked for various choices of $\chi$, and find monotonic trends in the behaviour of the recovered filament profiles. The results become more robust with more stringent cuts and we find convergence at the chosen value of $\chi=1/10$ for our choice of filament properties.

\begin{figure}[h]
    \centering
    \includegraphics[width=\textwidth]{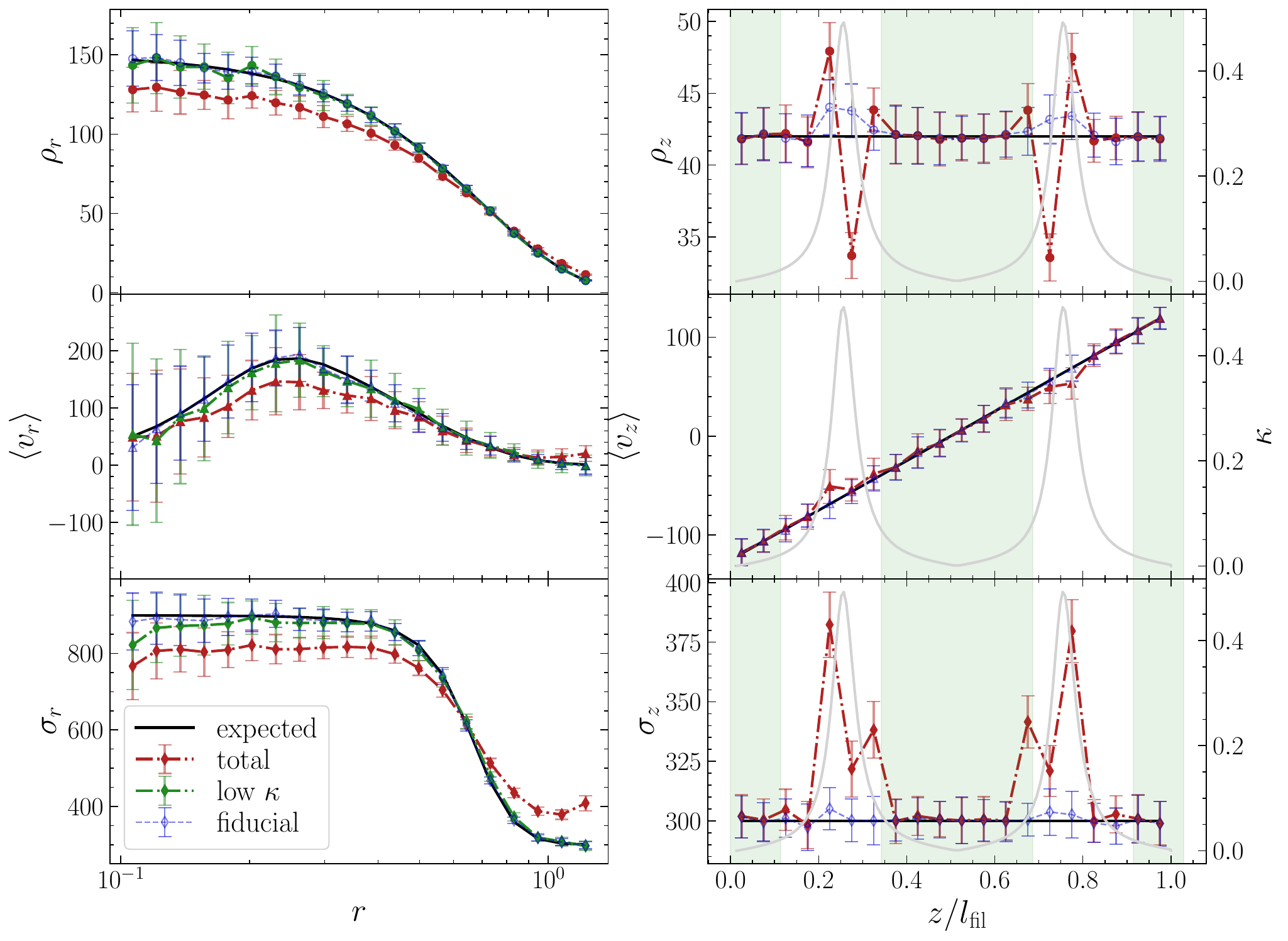}
    \caption{Effect of sampling of the spine on the density and velocity profiles. The panels are the same as in Figure~\ref{fig:curvature_split}, except the length along the spine ($z$) is scaled by the measured length $l_{\rm{fil}}$ of the discrete spine (sparsely and finely sampled for the corresponding cases). Black solid curves show the expected profiles, whereas blue open markers show the profiles obtained for the fiducial model. Solid red markers show the profile obtained for the fiducial filament, but when the number of points tracing the spine is reduced by a factor $5$. Green shaded regions in the right panel are the regions of low curvature, and the green markers in the left panels show the radial profiles calculated for only these regions. It is seen that sparse sampling of the spine induces biases in the inferred profiles, which can be alleviated by using only regions of low curvature. }
    \label{fig:ds_spine}
\end{figure}

\subsection{Effect of spine sampling}
\label{sec:spine_sampling}
The filaments in observations as well as simulations will be traced by discrete and finite tracers, and thus the filament spine can never be infinitely finely sampled. Even if one assumes that there is no error on the recovered spine (which is far from reality, and we shall explore this effect in the next section), the discreteness of the spine sampling can induce biases in the recovered profiles. The effect of sampling on the filaments recovered using the Bisous filament finder has been studied by \cite{Muru_tempel2021}. The effect of under-sampling will be more prominent when the tracers are massive halos (as opposed to less massive halos or dark matter particles), which are highly biased but few in number. Being strongly clustered, the more massive halos will tend to reside closer to the filament spines, which are regions of higher densities. Thus, the errors on the recovered spine are expected to be relatively low, but the effect of poor sampling of the spine would be prominent. Here, we quantify this effect and show that recovering the information from only low curvature regions can alleviate the biases, and the gain in accuracy is much more dominant over the loss of signal-to-noise ratio due to discarding high curvature regions.  

Here, we use the same filament configuration as in the fiducial case, the only difference being that the spine sampling is reduced by a factor $5$, keeping the sampling points equi-distant along the spine length. This spine is shown using blue dashed curves in Figure~\ref{fig:filament_illustration}. Note that in the case of biased tracers, there might be a local clustering along the spine (especially in high curvature regions), which we have not accounted for here. The red markers in Figure~\ref{fig:ds_spine} show the profiles recovered for the sparsely sampled spine, as opposed to the blue ones for the fiducial sampling. It is seen that the density in the inner regions is underestimated, and so is the mean radial infall. The radial dispersion profile is smoothed out. Also, both the density and tangential velocity dispersion estimates deviate strongly from the expectation in regions of high curvature. This is because although all the points sampling the spine lie exactly on top of the correct spine, the segments joining them deviate from the spine. This will be more prominent when the separation between the points is not much smaller than $R_\kappa$ at that point. Thus, the inferred radial distance as well as the radial direction of a region from the filament spine will have errors. Therefore, the regions that are considered for calculating the density very close to the spine will actually have contributions from outer regions of the filament, reducing the inferred density. Also, the radial vectors that are to be added while calculating velocity statistics will in reality have tangential components as well, thus smoothing out the statistics. 

In order to remove these biases, we split the spine based on curvature, placing the curvature threshold at $2/3$ of the inverse mean segment length. Again, the factor $2/3$ is open to choice, and we see monotonic trends in the recovered low curvature profiles when changing this factor, with convergence seen at the chosen value. We obtain the profiles only for the low curvature regions. This is shown using green markers in Figure~\ref{fig:ds_spine}, and the regions chosen are shown using green shaded areas in the right panels. There is a clear reduction of bias, with excellent recovery of the density profile and the outer regions of both velocity profiles. Some residual bias remains in the inner velocity dispersion profiles, which is nevertheless within the error bars. Thus, we have shown that splitting by curvature can lead to a more robust and unbiased recovery of profiles of filaments when the spine is sparsely sampled.
\begin{figure}[h]
    \centering
    \includegraphics[width=\textwidth]{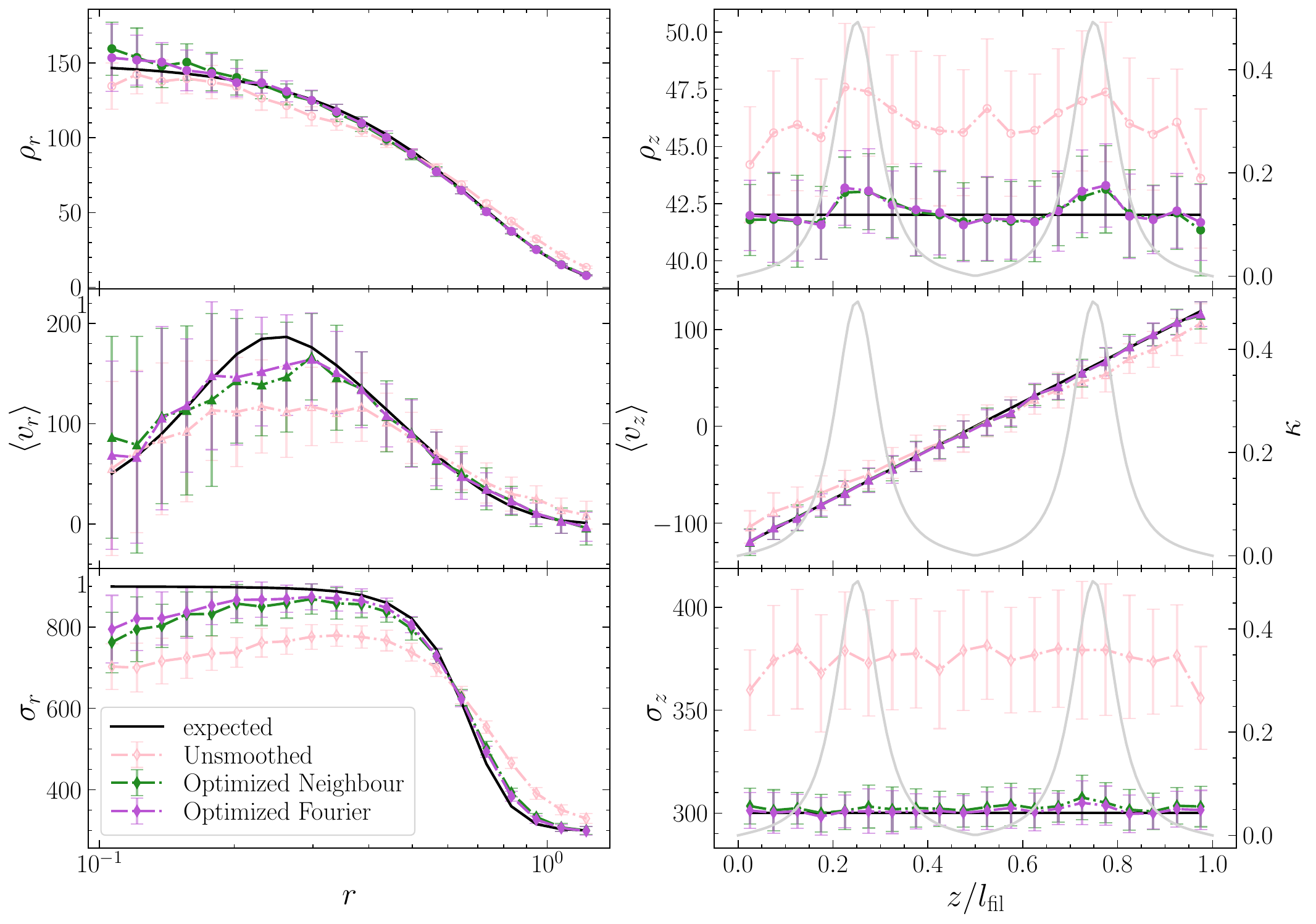}
    \caption{Effect of error in spine extraction on the estimated density and velocity profiles. The panels are the same as in Figure~\ref{fig:curvature_split}, except the length along the spine ($z$) is scaled by the measured length of the spine used  $l_{\rm{fil}}$. Black solid curves show the expected profiles, whereas open pink markers show the profile obtained for the fiducial filament, with a noisy spine. It is seen that strong biases are induced in the profiles due to the errors on the spine. Solid green (purple) markers show the profiles obtained using optimized neighbour (Fourier) smoothing, and retaining only low curvature regions for calculation of the radial profiles.  It is evident that proper smoothing of spines is necessary before predicting any profiles. See section~\ref{sec:error on the spine} for details.}
    \label{fig:noisy_spine}
\end{figure}

\begin{figure}[h]
    \centering
    \includegraphics[width=\textwidth]{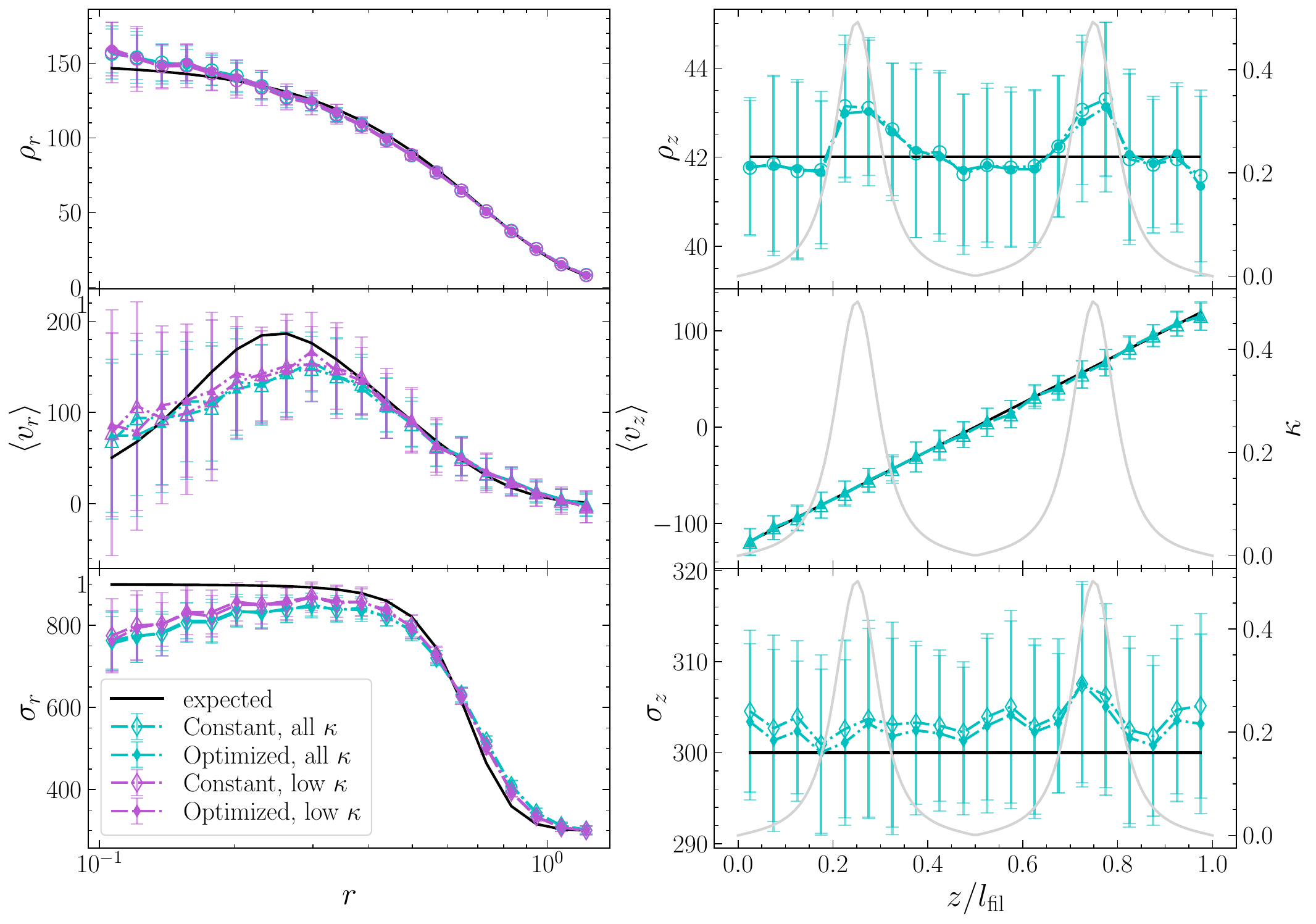}
    \caption{Comparison of optimized and constant smoothing and curvature segregation on the estimated profiles. The panels are the same as in Figure~\ref{fig:noisy_spine}. Black solid curves show the expected profiles, whereas coloured markers show the profiles obtained using neighbour smoothing. Solid (open) markers represent optimized (constant) smoothing. Cyan colour shows the radial profiles obtained for the whole filament, whereas purple colour shows those obtained for only the low curvature regions. It is seen that optimization does not provide much improvement when all the filaments belong to the same model.  See section~\ref{sec:error on the spine} for details.}
    \label{fig:NS_comparison}
\end{figure}

\begin{figure}
    \centering
    \includegraphics[width=0.7\textwidth]{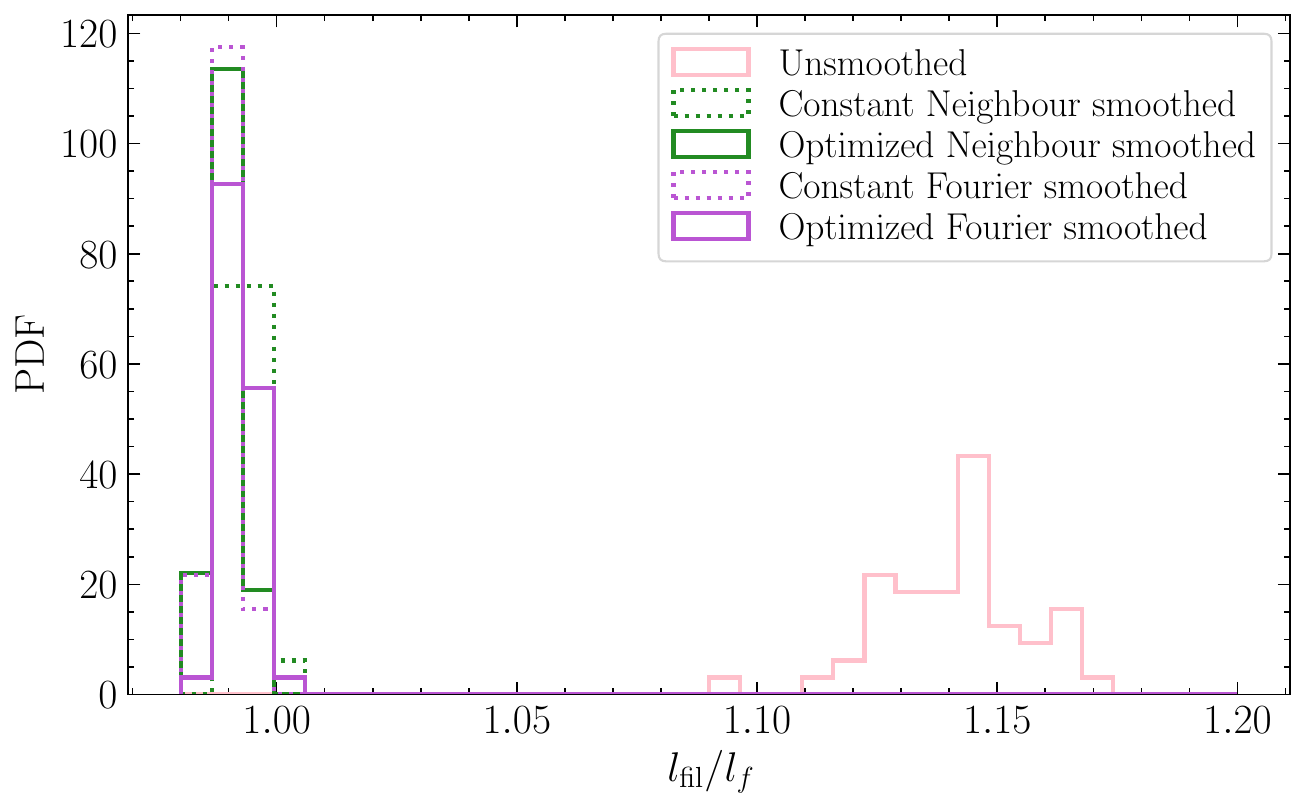}
    \caption{PDFs of lengths $l_{\rm{fil}}$ of the \disp\ inferred spines in units of the length $l_f$ of the actual spine. Pink colour represents the unsmoothed case, while green and purple colours represent the spines after applying a constant neighbour smoothing and optimized Fourier smoothing, respectively. Solid (dotted) lines show optimized (constant) smoothing.} We see that unprocessed spines lead to substantial errors on the lengths of the spines.
    \label{fig:length_distribution}
\end{figure}

\subsection{Effect of error on the spine}
\label{sec:error on the spine}
As explained in section~\ref{sec:spine processor}, there will always be errors on the spines recovered using filament finders, which will propagate into errors and biases in the estimated profiles. Here, we quantify this effect, and illustrate the improvements due to using an optimally smoothed spine. 

We use the same filaments as discussed in section~\ref{sec:fiducial}, but with spines recovered using \disp\ (see Appendix~\ref{app:filament spine identification} for details of spine recovery). The pink markers in Figure~\ref{fig:noisy_spine} show the profiles recovered using the noisy spines. Because of the noise, the inferred length of the spine is high, and one cannot compare the longitudinal profiles with the input profiles directly. Hence, the $x$-axis shows the length along the spine in units of the \emph{measured} length of the filament $l_{\rm{fil}}$  (which will always be longer than the true length $l_f$ in case of the noisy spine). We see that the deviation in the longitudinal profile is large not only in regions of high curvature (as seen before in the noise-free case), but now also in the low curvature regions. Thus, splitting by curvature alone will not help in this case. All the radial profiles also show strong biases. The density is over-estimated in the outer parts of the filaments, and is under-estimated in the inner parts. Both the mean and dispersion profiles of the radial velocity are smoothed out. These effects are sourced by mis-estimates, due to noise in the spine, of the distance from the true spine and, more importantly, the direction of the radial vector. 

As motivated in section~\ref{sec:spine processor}, smoothing can alleviate these biases. We smooth the filament spines by using two kinds of smoothings: (a) the widely used neighbour smoothing, and (b) Fourier smoothing (see Appendix~\ref{app:fourier_smo}, \ref{app:soothing_optimization} for details), and illustrate the results in Figure~\ref{fig:noisy_spine}. The smoothing parameters in both the cases have been optimized using our optimization technique, and only low curvature regions with $\kappa < 1/(10r_f)$ have been retained. The green markers show the profiles obtained by implementing the neighbour smoothing (see Appendix~\ref{app:neighbour_smo} for details). It is seen that this provides a huge improvement over the unsmoothed case, especially in the longitudinal profiles and radial density profiles. Fourier smoothing provides an equivalent performance. It is seen that except for the radial velocity dispersion ($\sigma_r$), all other profiles are in agreement with the expected profiles within the error bars. For $\sigma_r$, the deviations occur mainly close to the filament spine (small $r$). 
Slight deviations are to be expected, since one can never perfectly recover the exact spine, but the improvement with smoothing is encouraging.
Note that since here we are considering multiple realizations of a single filament model, our filament-by-filament optimization does not offer much advantage over constant smoothing, where the smoothing parameter is fixed by visual optimization of a single filament. To illustrate this, we smooth the spines using optimized and constant neighbour smoothing, and show the results in figure~\ref{fig:NS_comparison}, with and without discarding the high curvature regions. For constant smoothing, we have fixed $\Nsm=30$, which works well visually for this particular filament model (i.e., visual optimization). It is seen that profiles from optimized and constant smoothings are visually almost indistinguishable, whereas splitting by curvature provides a slight improvement (see the $\sigma_r$ profiles at large radii). One should note that even in the case of constant smoothing, some kind of optimization is still being performed; and the chosen value of $\Nsm$ is not arbitrary. Our technique prescribes a systematic way of optimization, with quantifiable justification. The importance of this kind of filament-by-filament optimization is more evident when the sample contains filaments with varied properties, as we demonstrate in section~\ref{sec:optimized_smoothing}.

To study the errors on the inferred lengths of the filament spines, the probability density function (PDF) of the lengths of the spines is calculated for the unsmoothed spines as well as processed spines using the two kinds of smoothings discussed above, with and without optimization. These PDFs are shown in Figure~\ref{fig:length_distribution}, where the lengths are given in units of $l_f$, the actual spine length. If there was no error on the spine, one would expect a Dirac-$\delta$ centred on unity. For our noisy spines, we see that the filament length is overestimated by $\gtrsim 10 \%$, with a wide spread. This error reduces to $\lesssim 2\%$ after smoothing. The width of the PDFs also reduce substantially. Here, where all the filaments belong to the same model, there is no preferred method of smoothing, and all the methods perform equally well. Thus, smoothing of the spines is essential even when one is dealing with statistics related to the lengths of the filament spines.

\section{Applications}
\label{sec:Applications}
Here, we discuss two applications of the methods and tools presented in this paper. The first is an application of optimized smoothing to a set of different filaments. The second is the effect of redshift space distortions (RSD) \cite{Kaiser87} on two simple cases: straight filaments aligned parallel and perpendicular to the line of sight (los).

\subsection{Optimized smoothing}
\label{sec:optimized_smoothing}

\begin{figure}
    \centering
    \includegraphics[width=\textwidth]{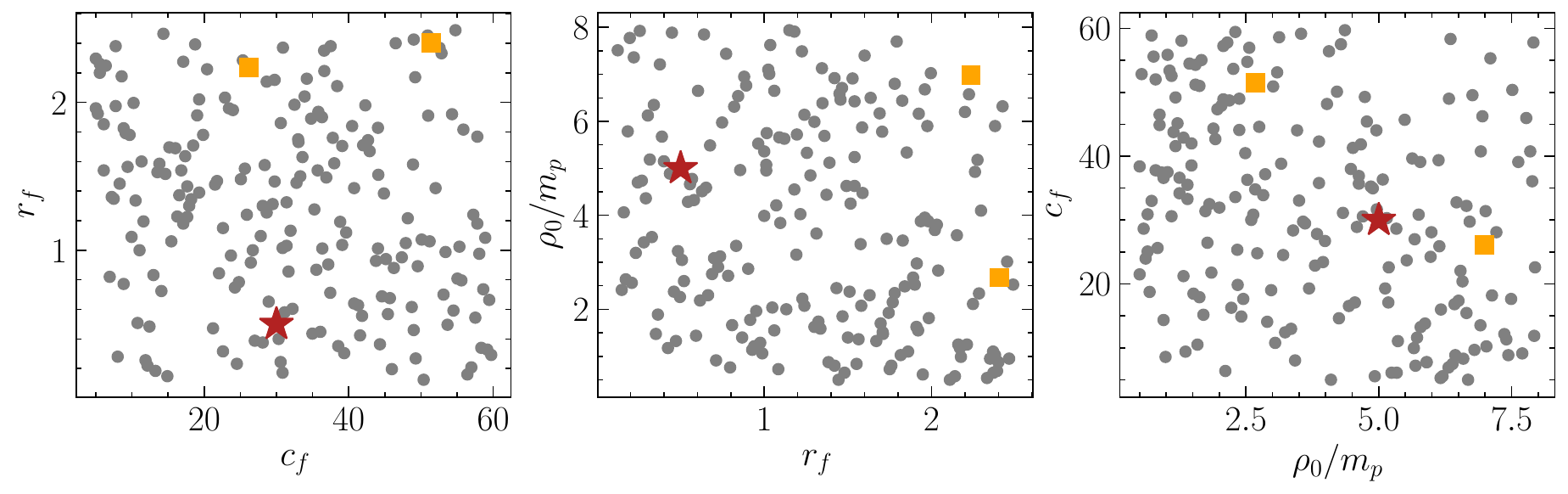}
    \caption{Distribution of the parameters of the filaments, $\{c_f, r_f, \rho_0/m_p \}$, used to generate a set of different filaments to study the application of optimized smoothing. The red star and orange squares represent the fiducial filament and the two filaments whose profiles are illustrated in Figure~\ref{fig:optimization_pdf} respectively (see section~\ref{sec:optimized_smoothing} text for details).}
    \label{fig:LHC_scatterplots}
\end{figure}

\begin{figure}
    \centering
    \includegraphics[width=\textwidth]{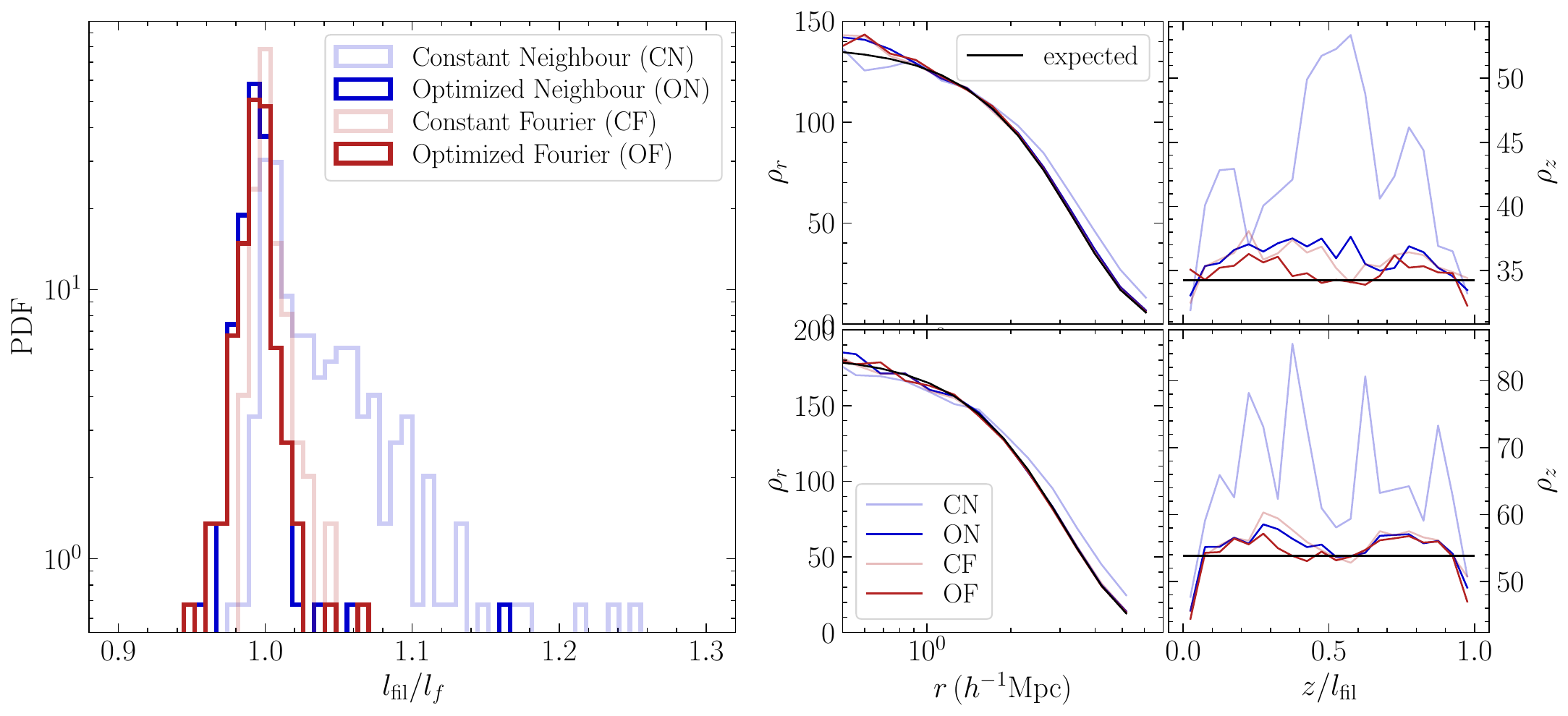}
    \caption{Illustration of need for optimized smoothing for a diverse filament set. The left panel shows the PDFs of filament spine lengths calculated over different filament models, after using various kinds of smoothings. The axes are the same as in Figure~\ref{fig:length_distribution}. Red (blue) colour represents Fourier (neighbour) smoothing, whereas dark (pale) lines show optimized (constant) smoothing. The four small panels on the right show the density profiles, each row corresponding to a different filament (see section~\ref{sec:optimized_smoothing} text for details). The colour scheme is the same as the one used in the left panel. We see that optimization gives excellent results for both neighbour and Fourier smoothing.}
    \label{fig:optimization_pdf}
\end{figure}

We have motivated the need for optimized smoothing in section~\ref{sec:spine processor}, and provided a detailed explanation in Appendix~\ref{app:soothing_optimization}. As seen in section~\ref{sec:error on the spine}, optimization does not provide much improvement over constant smoothing when all the filaments belong to the same filament model. However, this will not be the case in cosmological simulations. Here, we consider different models of filaments with the same $l_f$, and study the effect of multiple kinds of smoothings on their length PDFs. We generate $200$ filaments with the same sinusoidally curved input spine (see Figure~\ref{fig:smoothing_comparison} for visualization), but different filament parameters $(c_f, r_f)$  and different background number densities $\rho_0/m_p$. The distribution of these parameters is shown in Figure~\ref{fig:LHC_scatterplots}. The filament spines are identified using \disp\ (see Appendix~\ref{app:filament spine identification} for details).

The left panel in figure~\ref{fig:optimization_pdf} shows the length PDFs obtained for this set of $200$ filaments, smoothed using both optimized and constant, neighbour and Fourier smoothings. The constant smoothing parameters are fixed by visual optimization for the fiducial case, which correspond to $\Nsm=30, \, k_0=0.1 \,\hMpc.$ It is seen that the same constant neighbour smoothing does not work for all the filaments, giving rise to a wide spread in the inferred lengths $l_{\rm{fil}}${; whereas constant Fourier smoothing performs much better}. The spread substantially reduces as one shifts to optimized neighbour smoothing, with the peak of the PDF lying very close to unity, as is desirable. Optimized Fourier smoothing also gives excellent results.

To further study the effects of optimized smoothing, we focus on two particular filaments whose lengths are strongly overestimated in case of constant neighbour smoothing, with $l_{\rm{fil}}/l_f >1.2$. The parameters of these filaments are shown using orange squares in figure~\ref{fig:LHC_scatterplots}, whereas their radial and longitudinal density profiles are plotted in the four small panels in figure~\ref{fig:optimization_pdf}. The two rows represent the two filaments. In the case of radial profiles, we focus only on larger radii, where the error on the profiles is expected to be low. It is evident that constant neighbour smoothing leads to a strong bias in the recovered profiles, which reduces substantially with optimization. Comparatively, constant Fourier soothing fares much better, improving further with optimization. Overall, both neighbour and Fourier smoothing with optimization lead to excellent recovery of the radial density profiles, and are visually indistinguishable from the expected profiles in the figure.

Note that constant Fourier smoothing produces much better length PDFs as well as density profiles as compared to constant neighbour smoothing, as the noise always contributes towards high $k$ modes, which are cut off in Fourier smoothing. For this reason, we prefer and recommend Fourier smoothing over neighbour smoothing. Although both the optimized and constant Fourier smoothings produce comparable length PDFs and very similar density profiles, we recommend optimized smoothing over its constant counterpart (see Appendix~\ref{app:spine_smoothig} for details).

\begin{figure}[h]
    \centering
    \includegraphics[width=\textwidth]{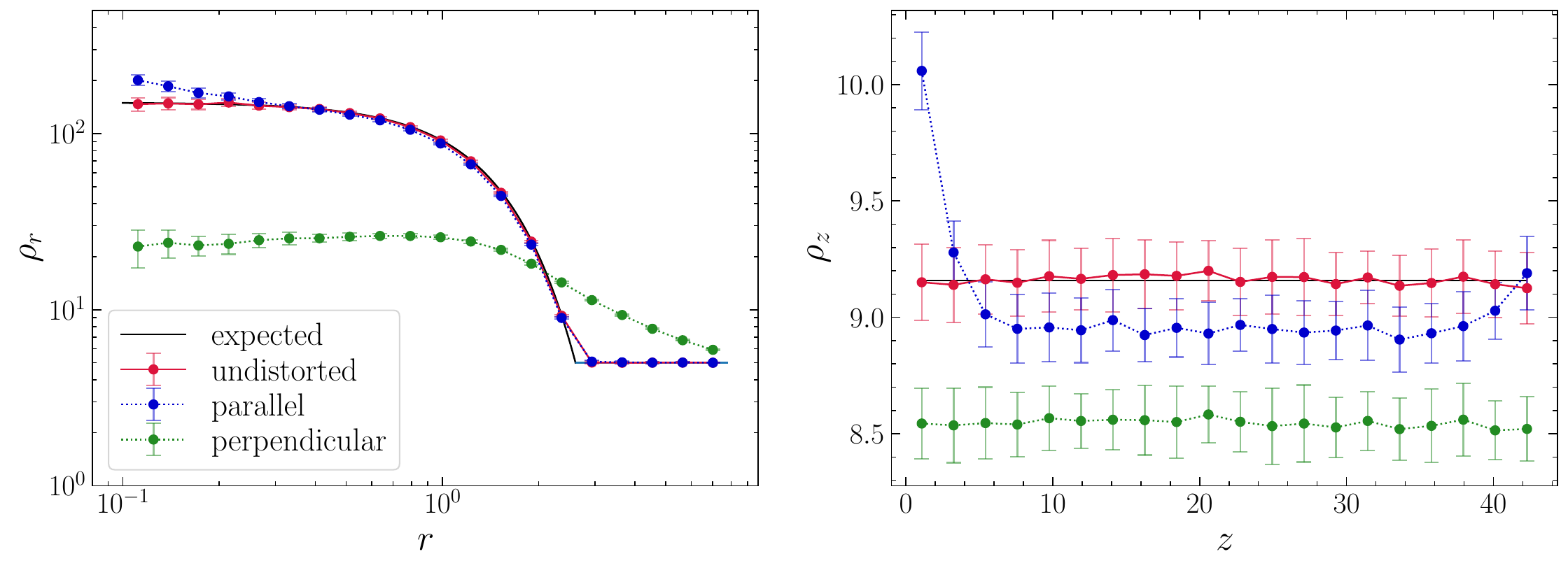}
    \caption{ Effect of redshift space distortions on the density profiles of straight filaments with nodes at both ends. The left (right) panel shows the radial (longitudinal) density profile. The black curves show the analytical expected profiles, whereas the red markers show the profiles obtained in real space.  The green and blue colours represent the redshift space distorted profiles when the line of sight is, respectively, perpendicular and parallel to the axis of the filament. The markers and error bars are obtained from the mean and standard deviation across $50$ realizations. See section~\ref{sec: RSD} for details.}
    \label{fig:rsd}
\end{figure}

\subsection{Redshift space distortions}
\label{sec: RSD}
Until now, we have focused on the recovery of profiles of filaments in real space. In real data, distances along the line of sight are inferred from observed redshifts which are affected by peculiar motions, leading to redshift space distortions (RSD). Here, we create simplistic but physically motivated velocity profiles for the filaments as well as the node halos and their environment, and study the effect of RSD on two simple cases: a straight filament aligned along and perpendicular to the line of sight, respectively. Since modelling of the velocity profiles is completely in our hands, we can switch off particular parts of the flows (e.g. dispersion in filaments, mean infall around halos, etc) and separately study the effect of various parameters on the RSD recovered filaments.

We consider an aperiodic box of side $80\, \Mpch$, and place a straight cylindrical filament of length $l_f=50 \, \Mpch$ and radius $r_f=1 \, \Mpch$. The filament concentration is $c_f=30$, and its velocity profiles are the same as those used for the fiducial filament. An NFW halo is placed at each end of the filament; the radii ($R_{200c}$) and concentrations of the two halos are $R_1=0.6 \, \Mpch, \, R_2=0. 4 \, \Mpch$ and $c_1=5, \, c_2=8$ respectively. The halo outskirts are chosen to have a power law density profile $\rho_{2h} \propto \rsp^{-3}$, and the normalization is chosen such that the densities of the NFW halo and the halo outskirts match at the halo boundary. The velocity dispersion in the halo outskirts is assumed to be isotropic, and matches that of the background, with 1D velocity dispersion of $300\, \kms$. The mean tangential velocity in the halo outskirts is assumed to vanish, and the mean radial infall is modelled using $\langle v_r \rangle =-D/ \chi$, with $D_1=500 \, \kms, \, D_2=400\, \kms$, and $\chi$ is the value of $\rsp$ in units of radius of the respective halo. Although in reality, the halo outskirts will not be radially symmetric and simple, here our main goal is just to illustrate qualitatively how \filgen\ and \filapt\ can be used to study RSD in filaments, and the choice of profiles is not very important. The filament profiles are estimated by taking a spine that starts and ends at the outer boundary of the halo outskirts, so that the halos and their outskirts would be excluded from the profiles in real space. 

The straight filament axis is taken to be along the $z$-direction. We consider two cases, one where the line of sight is along the $z$-axis and the other where it is along the $x$-axis. Let the Cartesian coordinate along the line of sight be represented by $\alpha$ ($\alpha=z, x$ in the two cases, respectively). The positions of the particles along the line of sight will get distorted, and are given by:
\begin{equation}
\label{eq:RSD}
    \alpha_{\rm{RSD}}= \alpha + \frac{v_\alpha}{aH} \,,
\end{equation}
where $a, H$ are the scale factor and Hubble parameter respectively.\footnote{For this demonstration, we work at the current epoch ($a=1$), so that $H=H_0=100h\,\kms{\rm Mpc}^{-1}$, the Hubble constant. Since our units of length are \Mpch, the value of the Hubble constant is irrelevant.} We shift the positions of all the particles according to this equation, and calculate the profiles, excluding regions up to $4R_{200m}$ on either side of the spine corresponding to the respective nodes and their outskirts (c.f., the discussion in section~\ref{sec:tools}). The profiles are calculated by taking a finely sampled set of points on the input spine of the filament; we do not consider the noisy spines obtained by applying a filament finder. We only show the radial and longitudinal density profiles in Figure~\ref{fig:rsd}. Here, we have probed radii well beyond the boundary of the filament $R_f$, since the filament after being distorted can get spread out.

First, looking at the case where the line of sight is perpendicular to the filament axis (green markers), one sees that both the longitudinal and inner radial density is lowered, whereas the outer radial density has increased. Two effects will play opposing roles in this case. The radial infall will tend to compress the filament along the line of sight increasing the density, whereas the radial dispersion will tend to spread it out (similar to the usual Fingers-of-God (FoG) effect), decreasing the density. For our choice of parameters, the second (FoG smearing) effect is dominant. This leads to a decrease in longitudinal density, which is calculated by averaging over a disc of fixed radius $R_f$. FoG also smears the radial profile, increasing the density in filament outskirts, and decreasing it inside.

Second, considering the case where the line of sight is along the axis of the filament (blue markers), many effects contribute, leading to complicated changes in the filament profiles. The halo outskirts have a strong radial infall, hence these particles will predominantly be shifted towards the nodes, not contributing to any changes in the filament profiles. The strong velocity dispersion of the nodes, however, will spread them out (the usual FoG effect), substantially increasing the density in the filament, especially at its edges, 
which is seen in the longitudinal profile. The asymmetry in the increase is because one of the nodes is more massive than the other. This effect also leads to an increase in the density close to the axis of the filament, as seen from the radial profile. The radial infall on top of the filament is perpendicular to the line of sight, and thus will not contribute to any changes in redshift space. Finally, the mean flow of the particles away from the filament centre towards the nodes causes a decrease in density in the central part of the filament, as seen from the longitudinal plot. 

One has to note that the RSD effects will vary as one changes the density and velocity profiles of the filaments, as well as nodes and their outskirts. Here, we have only illustrated that \filgen\ and \filapt\ can in principle be used to study the filaments in redshift space. We leave a detailed discussion of redshift space applications for future work.

\section{Conclusion}
\label{sec:conclusion}
Cosmic filaments are visually the most striking features of the cosmic web. Unlike halos and voids, however, their potential in helping understand galaxy evolution or cosmology remains largely untapped. It is not clear, e.g., whether filaments exhibit universality in their density and/or velocity profiles similar to that seen for halos and voids. In part, this is due to a lack of consensus on what exactly constitutes a filament. For example, multiple filament finders exist in the literature, and each one identifies rather different filaments for the \emph{same} data set \cite{Libeskind_et_al2018}. In the absence of a notion of a `true' spine, it is then difficult to objectively choose a filament finder and robustly fix its parameters; the general practice is to rely on visual inspection. 

In this work, we have taken steps towards rectifying this lack of objectivity in the identification and analysis of cosmic filaments, by introducing two tools. The first -- the {\bf Fil}ament {\bf Gen}erator, or \filgen\ -- generates the complete phase space information of particles in realizations of mock filaments with a \emph{known} spine and some input density and velocity properties, all of which can be made arbitrarily realistic/complex (see section~\ref{sec:filgen}). Operationally, \filgen\ first samples the user-specified density and velocity profiles assuming a straight filament, and then self-consistently curves the filament according to a user-specified spine curve. Thus, \filgen\ synthesizes particle realizations of the `truth', solving the problem of arbitrariness of judgement when subsequently using a filament finder. For simplicity, in this work, we have focused on generating filaments whose radial density profiles can be described by a single Gaussian as a function of perpendicular distance $r$ from its spine.

Although knowing the `truth' regarding the filament underlying a given particle data set is important, it is equally important to identify robust statistics that maximise the information retained after applying a filament finder. E.g., it is unlikely that the `true' radial or longitudinal density and velocity profiles of a filament would be accurately recovered if the spine identified by the filament finder is very noisy. This issue is addressed by our second tool -- the {\bf Fil}ament {\bf A}nalysis and {\bf P}rocessing {\bf T}ool, or \filapt\ -- which provides three main post-processing modules for use on the data obtained from a filament finder (see section~\ref{sec:filapt}): \begin{itemize}
\item \filapt\texttt{.ExtractProfiles} extracts the profiles of density along with mean and dispersion statistics for different velocity components, as a function of perpendicular distance from the spine or length along the spine. This module inherently assumes that the spine being used is `perfect', thus allowing its output to be used in various quality tests in combination with \filgen.
\item \filapt\texttt{.SmoothSpine} \emph{optimally} smooths each filament spine identified by the filament finder. The optimization uses a novel method which minimizes the estimated filament radius, for reasons discussed in section~\ref{sec:spine processor}. Not all filaments in the cosmic web are the same, leading to different error properties on the individual inferred spines, and we have demonstrated that optimization is therefore essential. E.g., filament lengths can be substantially overestimated in the absence of optimization (Figure~\ref{fig:optimization_pdf}). Based on the results of tests in section~\ref{sec:optimized_smoothing}, we also recommend smoothing in Fourier space, rather than the commonly used neighbour smoothing.
\item \filapt\texttt{.ComputeCurvature} estimates the local curvature $\kappa$ along the spine (see Appendix~\ref{app:curvature}). \filapt\ further provides for each filament to be split by $\kappa$, thus revealing the curvature dependence of the extracted profiles. We have shown in section~\ref{sec:spine_sampling} that regions with the highest $\kappa$ along any filament are the major contributors of biases in the extracted radial profiles, and discarding these can substantially clean the profiles. The definition of ``high $\kappa$'' is not completely unambiguous, and the value has to be decided taking into consideration the radius of the filament as well as the mean length of the segments of the filament spine. In general, rather than discarding the high curvature profiles, one would like to study them separately. To our knowledge, this is the first recognition that selecting on spine curvature may be critical in obtaining unbiased filament demographics.
\end{itemize}

As a simple initial application of \filgen\ and \filapt, we have shown how redshift space distortions can potentially bias the inference of the underlying filament properties (section~\ref{sec: RSD}). Ultimately, we are interested in uncovering universality in the phase space profiles of filaments in the cosmic web. For this purpose, one would like to first work in real-space dark matter filaments in simulations. In this case, the biases will arise primarily from errors on the inferred spines, spine sampling and curvature. We have discussed all these biases and ways to alleviate them in this paper. We also briefly mention a more detailed application using N-body simulations that we are currently pursuing, at the end of the section. 

There are, however, several potential improvements possible in \filapt\ and \filgen. We list these caveats here and will address them in future work.
\begin{itemize}
\item As mentioned at the end of section~\ref{sec:effect of radius}, the radial profile produced by \filgen\ in regions when the local radius of curvature $\kappa^{-1}$ is comparable to or smaller than the filament thickness $r_f$, does not match the input profile. If the goal is to generate a specific radial profile in regions of high curvature, an iterative procedure would be necessary, where the
azimuthal profile of the straight filament is changed iteratively so that after curving of the spine, the radial profile corresponds to the required one. 
\item \filgen\ currently does not model the multi-scale nature of filaments; the presence of substructure (halos in filaments, `sub-filaments') is not accounted for. Similarly, \filapt\ currently assumes each filament to be a monolithic structure.
\item Filaments traced by biased tracers such as halos or galaxies are not yet modelled by \filgen. In principle, it should be possible to include density fluctuations and use excursion-set inspired biased point processes such as density peaks \cite{ps12} as proxies for halos/galaxies.
\item The halo outskirts, relevant for modelling and understanding redshift space effects, are currently not modelled well by \filgen. This is primarily due to a lack of analytical understanding of the so-called 1-halo to 2-halo transition regime. It should, however, be possible to mitigate this issue using controlled calibrations of halo surroundings in N-body simulations.
\item \filapt\texttt{.SmoothSpine} currently optimizes the smoothing parameters (e.g., the cutoff wave number in case of Fourier smoothing) for each filament assuming that the radial density profile is well described by a \emph{single} Gaussian (as is currently modelled by \filgen). Filaments in simulations are typically more complex \cite{wang+24, 2pop2020}, and we are currently exploring the performance of Gaussian mixture models for the optimization step.
\end{itemize}

Taken together, \filgen\ and \filapt\ can be used to calibrate any given parametrised filament finder before applying it to simulation data. We will report the results of such an exercise using \disp\ \cite{Disperse_theory,Disperse_illustration} in the near future. This would enable a systematic and simultaneous exploration of multiple filament properties including length, thickness, mean overdensity, concentration, velocity anisotropy, etc., in N-body simulations, with the ultimate aim of uncovering possible universality in filament demographics. An even more exciting prospect is to be able to robustly distinguish between filament properties for different dark matter models \cite[e.g.,][]{aragon-calvo24,bl24}. We will take up this programme in future work.

\section*{Data availability}
The code for implementing \filgen\ and \filapt\ is publicly available at \url{https://github.com/SaeeDhawalikar/Filtools}. 

\acknowledgments

We thank the anonymous referee for a very constructive report.
The research of AP is supported by the Associates Scheme of ICTP, Trieste. This work made extensive use of the open source computing packages NumPy \citep{vanderwalt-numpy},\footnote{\url{http://www.numpy.org}} SciPy \citep{scipy},\footnote{\url{http://www.scipy.org}} Pandas \citep{mckinney-proc-scipy-2010, reback2020pandas},\footnote{\url{https://pandas.pydata.org}} Matplotlib \citep{hunter07_matplotlib},\footnote{\url{https://matplotlib.org/}} and Jupyter Notebook.\footnote{\url{https://jupyter.org}} The analysis was performed on the Pegasus cluster at IUCAA, Pune.\footnote{\url{http://hpc.iucaa.in}}

\bibliography{references}

\appendix

\section{Curvature $\kappa$ of a curve}
\label{app:curvature}
Here, we first give the standard mathematical definition of curvature $\kappa$ for a continuous curve. Based on this definition, we give a formula to calculate $\kappa$ for a discretely sampled curve, and show that this is in good agreement with the analytical definition.

\subsection{Continuous curves}
\label{app:curvature_continuous}

Curvature quantifies how much a given curve deviates from a straight line, or by how much the tangent to the curve changes its direction per unit length along the curve. In calculus, curvature $\kappa$ for a curve $\mathcal{C}$ is defined to be \cite[Chapter~14]{apostolcalculus}:
\begin{equation}
\label{eq:kappa_basic}
    \kappa \equiv \left\lVert \frac{ d \hat{T}}{dl_s}\right\rVert \,,
\end{equation}
where $\hat{T}$ is the unit tangent, and $l_s$ is the length along the curve. For a general parametrization of the curve $\mathcal{C} = \vec{C(}t)$, the curvature can be written as:
\begin{equation}
\label{eq:kappa_parametric}
    \kappa (t)= \frac{\lVert \hat{T}'(t) \rVert}{\lVert \vec{C}'(t)\rVert}= \frac{\lVert \vec{C}''(t) \times \vec{C}'(t)\rVert}{\lVert \vec{C}'(t) \rVert^3}\,,
\end{equation}
where $' \equiv d/dt,\, '' \equiv d^2/dt^2$, and the second equality follows from using chain rule and the fact that $\hat{T} \perp \hat{T}'$ (\cite{apostolcalculus}, equation~14.22). It is seen from equation~\eqref{eq:kappa_basic} that $\kappa$ has units of inverse length. For a circle, $\kappa$ is constant, and is equal to the inverse of its radius, whereas it is identically zero for a straight line. For an arbitrary curve, the inverse of $\kappa$ at a given point is called the radius of curvature $R_\kappa=\kappa^{-1}$.

\subsection{Discretely sampled curves}
\label{app:curvature_discrete}
\begin{figure}[h!]
    \centering
    \includegraphics[width=0.7\linewidth]{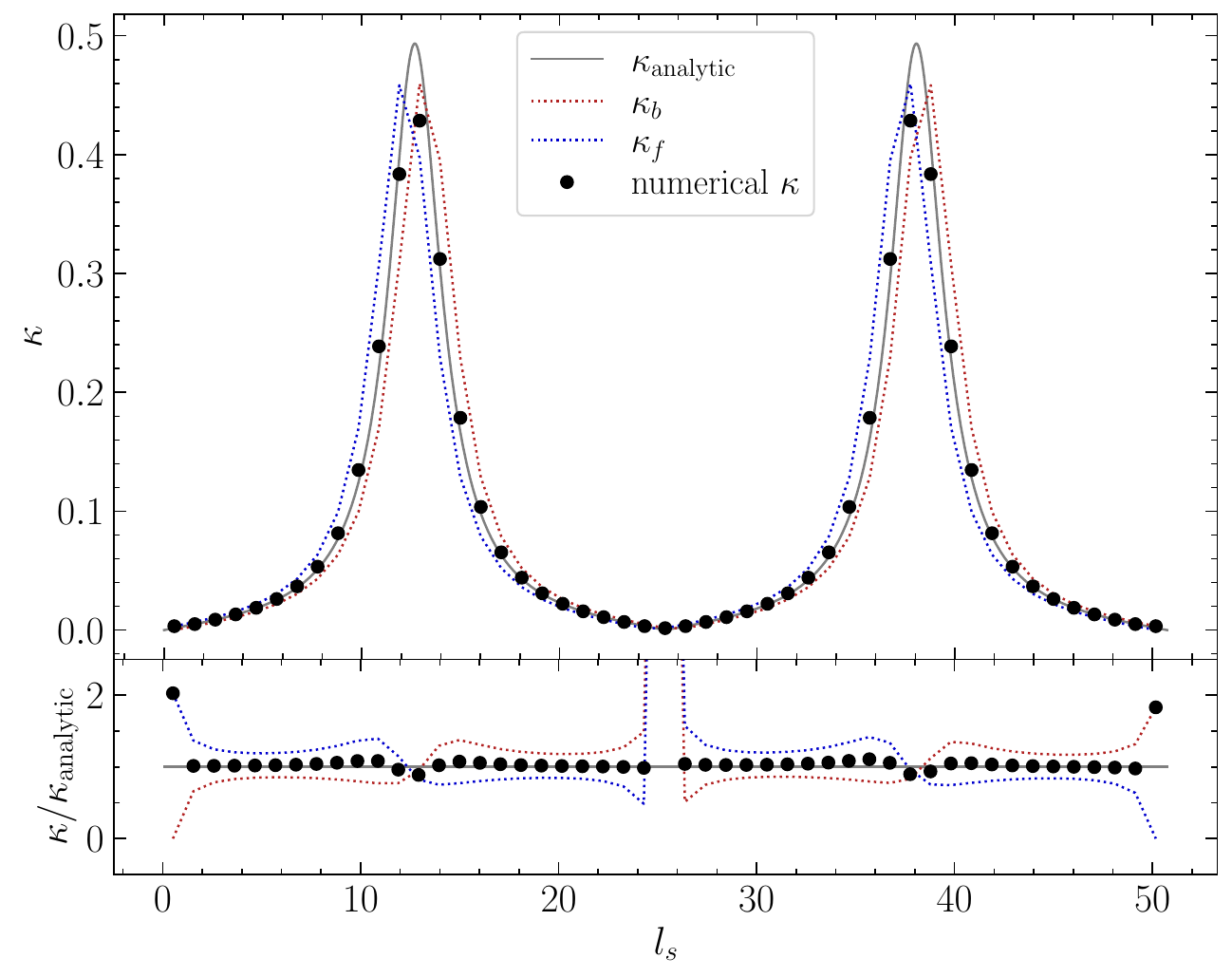}
    \caption{Comparison between the analytical curvature of a curve and the one obtained using \eqns{eq:kappa_f}-\eqref{eq:kappa_final}. The curve used is the spine of the filament described in section~\ref{sec:fiducial}. In the top panel, the curvature $\kappa$ is plotted as a function of length along the curve $l_s$. Grey solid curve shows the analytical value $\kappa_{\rm analytic}$. Blue and red dotted curves show the curvature calculated for the discretely sampled curve using the directions of outgoing and incoming segments, respectively. Black markers show the final curvature values calculated using equation~\eqref{eq:kappa_final}. 
    Bottom panel shows a ratio between the calculated curvature and $\kappa_{\rm{analytic}}$. The ratio blows up near the centre because the denominator is approaching zero. We see that there is an excellent agreement between the black points and the analytical curvature.}
    \label{fig:curvature}
\end{figure}
We have introduced the standard definition of curvature above, which defines $\kappa$ for a continuously differentiable curve parametrized by a single parameter $t$. However, we want to find the curvature of filaments found by filament finders, which usually define the filament spine to be a set of points rather than an analytical curve. Here, we define the curvature of each segment of the spine based on the basic definition that $\kappa$ is the rate of change of the angle of the tangent per unit length (equation~\ref{eq:kappa_basic}): \begin{equation}
\label{eq:kappa_i}
    \kappa_i \equiv \frac{\lvert \cos^{-1}(\vec{p}_{1\,i} \cdot \vec{p}_{2\,i})\rvert}{l_i}\,,
\end{equation}
where the subscript $i$ refers to a particular segment, $l_i$ is the length of the segment, and $\vec{p}_{1,2\,i}$ are the unit vectors specifying the direction of the tangent to the curve at the start and end points of the segment respectively. Now, the direction of the tangent at a point in this discretely sampled curve can either be taken to be the direction of the incoming segment at the point, or that of the outgoing segment. Thus, one can define two curvature values, corresponding to the two choices of $\vec{p}_{1,2\,i}$ used.  Let the $i$th segment join the points $\vec{r}_i,\, \vec{r}_{i+1}$, with $\vec{l}_i \equiv \vec{r}_{i+1}-\vec{r}_i$ and $l_i \equiv \lVert \vec{l}_i \rVert$. Then the two curvature values for this segment are given by:
\begin{align}
    \kappa_{f,i} & \equiv \frac{1}{l_i}{\left| \cos^{-1} \left(\frac{\vec{l}_i \cdot \vec{l}_{i+1}}{l_i \, l_{i+1}} \right)\right|}\,,
    \label{eq:kappa_f}\\
    \kappa_{b,i} & \equiv \frac{1}{l_i}{\left| \cos^{-1} \left(\frac{\vec{l}_{i-1} \cdot \vec{l}_{i}}{l_i \, l_{i-1}} \right)\right|}\,,
    \label{eq:kappa_b}
\end{align}
Here, $\kappa_f, \kappa_b$ are the curvatures calculated using the directions of the outgoing (or `forward') and incoming (or `backward') segments, respectively. We define the final curvature value for the segment to be the average of these two quantities, accounting for edge effects at the ends of the filament:
\begin{equation}
\label{eq:kappa_final}
    \kappa_i \equiv 
     \begin{cases}
       (\kappa_{f,i}+\kappa_{b,i})/{2} &\,; \quad 1< i < N\\
       \kappa_{f,i} &\,; \quad i=1\\
       \kappa_{b,i} &\,; \quad i=N\\
     \end{cases}
\end{equation}
where $N$ is the total number of segments in the given filament. Figure~\ref{fig:curvature} shows a plot of $\kappa_f, \kappa_b$ and the average $\kappa$ given in the above equation along with the analytical expectation for a curve. The final values of $\kappa$ calculated using our approach (represented using black markers) are in good agreement with the analytical values, showing that this is a valid and reliable method to calculate curvatures of discretely sampled curves.

\section{Smoothing the filament spine}
\label{app:spine_smoothig}
Here, we briefly describe two of the smoothing techniques used to process the spines obtained using a filament finder: the widely used neighbour smoothing (which is usually used to process \disp\, spines), and Fourier smoothing, which we introduce in this paper. We also describe in detail a way to optimize the smoothing parameters for any given filament.

\begin{figure}
    \centering
    \includegraphics[width=\textwidth]{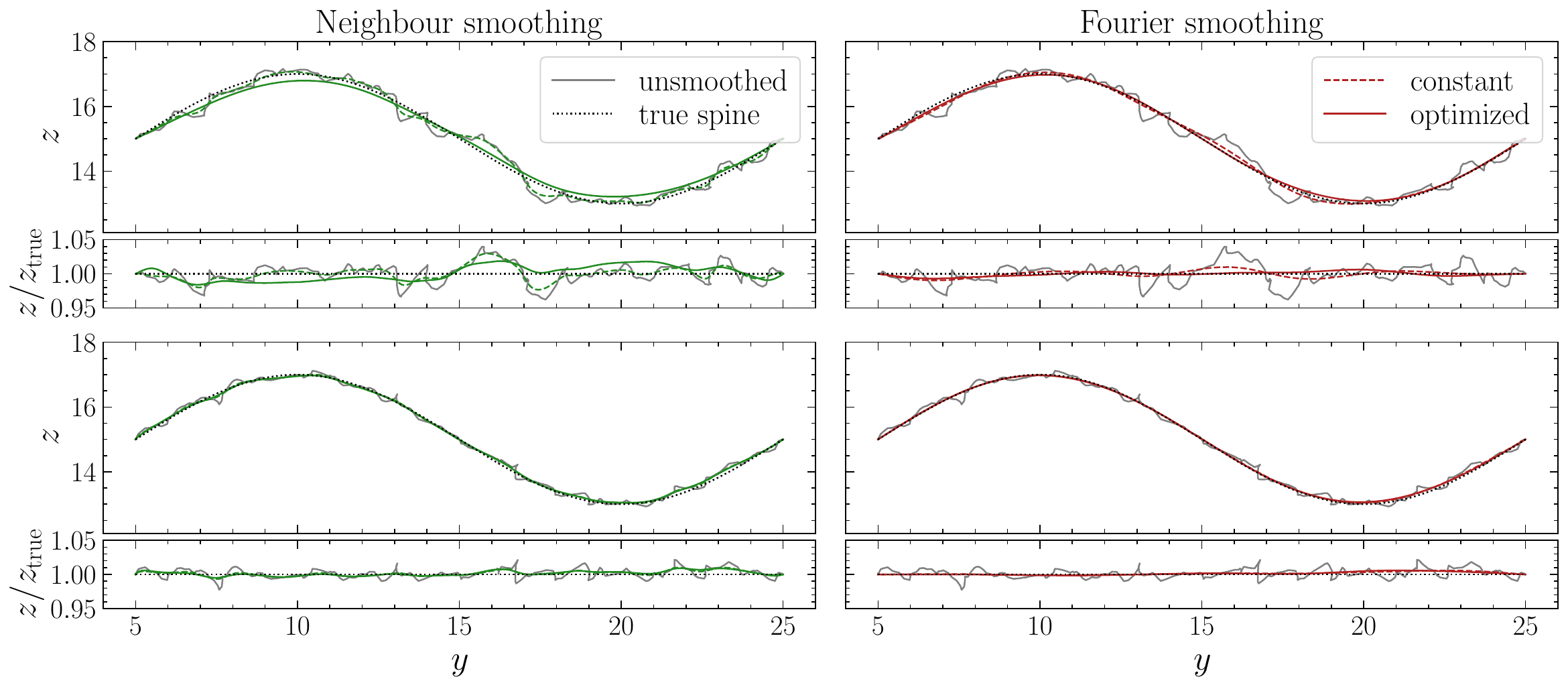}
    \caption{Comparison between different kinds of smoothings applied to two filaments with the same input spine and similar $c_f, \rho_0$, but different $r_f$. The main panels show the $x$-projections of the spines of two filaments. In the narrow panels below each main panel, the $z$-coordinate is divided by the corresponding $z$-coordinate of the true spine, to better illustrate the deviations. The filament in the top panels has approximately twice the $r_f$ as compared to the one shown in the lower panels. The black dotted curves represent the input spine, whereas the unsmoothed \disp\ spine is shown using solid grey curves. Left (right) panels show neighbour (Fourier) smoothing. In each case, optimized and constant smoothings are represented using solid and dashed curves, respectively. We see that the optimized Fourier smoothing gives excellent results in recovering the spines of both the filaments.}
    \label{fig:smoothing_comparison}
\end{figure}

\subsection{Neighbour smoothing}
\label{app:neighbour_smo}
In the neighbour smoothing, the end points of the filament are fixed, and the smoothed positions of all the other points are given by:
\begin{equation}
\label{eq:neighbour_smoothing}
\vec{r}_{i, \, s}= \frac{1}{4}\left(\vec{r}_{i-1}+2\vec{r}_i+\vec{r}_{i+1} \right) \,
\end{equation}
where $\vec{r}_i$ is the position of the point to be smoothed, and $\vec{r}_{i \pm 1}$ are those of its immediate neighbours. The smoothing is performed \Nsm\ number of times to give the smoothed filament.

The green dashed curves in the first column in Figure~\ref{fig:smoothing_comparison} show the application of this smoothing to two filaments similar in all regards except for $r_f$, which differs by a factor $\approx 2$ (see the Figure caption for details), for $\Nsm=30$. We see that this choice of smoothing works for the filament with smaller $r_f$ (bottom two panels), but the thick filament (top two panels) remains under-smoothed. 

\subsection{Fourier smoothing}
\label{app:fourier_smo}
The idea here is to smooth the filaments in Fourier space using a low-pass filter. This is a valid approach, as the noise will contribute to high wave numbers, and removing these should lead to recovery of the actual spine. First, we calculate the length along the spine $l_s$, so that the curve (or spine) is parameterized by this length. Then, we interpolate to get equi-spaced points on the spine. The three Cartesian coordinates of all the points tracing the curve are separately Fourier transformed with respect to $l_s$. A rounded top hat \cite{Bond_et_al1991} low pass filter (RTHF) with the same parameters is applied to all three Fourier spectra. The RTHF is defined as a convolution of a sharp-$k$ low pass filter and a Gaussian in Fourier space:
\begin{equation}
\label{eq:RTHF}
    {\rm RTHF}(k, k_0, q) = \frac{1}{2}\left[ \erf{\frac{k+k_0}{\sqrt{2q^2}}}- \erf{\frac{k-k_0}{\sqrt{2q^2}}}\right]\,.
\end{equation}
Here, $k$ is the Fourier space variable, $k_0$ is the cutoff wave number of the low-pass filter, $q$ is the standard deviation of the apodizing Gaussian, and $\erf{x}$ is the error function. We do not use a sharp-$k$ filter, as it leads to ringing in real space. We have checked that the smoothed spine is relatively independent of the choice of $q$, and a value $q=k_0/5$ works well enough for any kind of spine or $k_0$. Thus, the only free parameter is the cutoff wave number $k_0$. There are undesirable edge effects if one directly uses the Fourier transform of the input spine for smoothing. Instead, we first pad the spine on either side with an odd padding of length equal to the length of the spine. This padded curve is Fourier transformed, the filter is applied, and an inverse Fourier transform is performed to get back to the real space. This curve is cut off on either side to remove the initial padding, and get the smoothed spine. We choose odd padding as it fixes the end points. However, one can also opt for constant or even padding, and only small differences are seen at the end points of the smoothed filament, those too when $q$ is very low.

The second column in Figure~\ref{fig:smoothing_comparison} shows Fourier smoothing applied to two different filaments. We have chosen a constant $k_0=0.1\,\hMpc ; \, q=k_0/5$, and the smoothed spines are shown using dashed red curves. This choice of parameters works well for the thinner filament (bottom two panels). Although the smoothed spine of the thicker filament (top two panels) shows some deviations, the performance of constant Fourier smoothing for the thick filament is much better than the constant neighbour smoothing. This is because although the noise in this case has higher amplitude, it still contributes to higher wave numbers in Fourier space, which are killed using the low pass filter. Another advantage of Fourier space smoothing is that the cutoff wave number is bounded; the mean spacing between the points in the input spine provides an upper bound on $k_0$, and the length of the spine provides a lower bound. There is no such upper bound on \Nsm\ in neighbour smoothing. Although Fourier smoothing offers an improvement, some longer wavelength noise modes are retained, and the choice of smoothing parameters is still subjective. We address this issue in the next section.

\subsection{Smoothing optimization}
\label{app:soothing_optimization}
As seen from Figure~\ref{fig:smoothing_comparison}, different types of filaments lead to different noise properties of the spines obtained using a filament finder (in this case, \disp). Thus, a particular choice of smoothing that works for one filament might not work for another; and there is a need to optimize the smoothing parameters separately for individual filaments. Here, we introduce a physically motivated way to achieve this. 

One expects the filaments to be densest at their core, and the density is supposed to drop with increasing $r$, to reach a constant value at large $r$. This constant value might not be equal to the mean density $\rho_0$, since the filaments could be embedded in different environments like a sheet or a void. When the spine of the filament is not robustly estimated (under- or over-smoothed), the radial density profile of the filament will get smoothed out (inner, denser regions contributing to larger $r$ while estimating the profiles and vice versa), increasing the width of the density profile. Therefore, it is logical to assume that the choice of smoothing parameters that gives rise to the narrowest radial density profile is the optimum choice. 

Here, we use filaments generated using \filgen, which have (by construction) a Gaussian radial density profile. Thus, there is a unique way to estimate the width of the profile, which is the standard deviation of the Gaussian, and is equal to $r_f$. For any given choice of smoothing, the smoothed spine is used to estimate the radial density profile, and the errors are taken to be Poissonian. The profile is fitted with a Gaussian (see \eqn{eq:Gaussian}), with two free parameters $(c_f,  r_f)$. $\rho_0$ is fixed, as it is completely degenerate with $c_f$. Note that here, we choose to fit the complete, un-truncated Gaussian. This procedure is repeated for different values of smoothings, till a converged minimum $r_f$ is reached. This corresponds to the optimum smoothing. We use the publicly available code \texttt{PICASA} \cite{picasa} to implement anisotropic simulated annealing (ASA), which is a fast and robust method to achieve the optimum. In future work, we will extend the technique so as to estimate the filament thickness using more flexible functions than a single Gaussian.

The results of implementing this optimization using neighbour (Fourier) smoothing are illustrated in the left (right) panels of Figure~\ref{fig:smoothing_comparison}, using solid green (red) curves. There are no visual differences between the constant and optimized smoothing for the thin filament (lower panels). For the thick filament (upper panels), optimized neighbour smoothing does provide an improvement. Although the smoothed spine does not coincide with the input spine completely, a substantial amount of the noise has been removed, and the smoothed spine agrees better with the input as compared to the case of constant neighbour smoothing. We see that optimized Fourier smoothing gives an excellent performance in recovering the spines of both the filaments. Overall, it is evident that optimization (especially using Fourier smoothing) gives a substantial improvement over the widely used constant neighbour smoothing, along with eliminating the ambiguity associated with the smoothing parameters.

\section{\disp\ filament identification}
\label{app:filament spine identification}
To obtain the spines of the filaments explained in sections~\ref{sec:fiducial} and \ref{sec:optimized_smoothing}, we first add two NFW halos at the ends of these filaments, with radii $R_{200c}=0.4, \, 0.6\, \Mpch$ and concentrations $c=8, 5$ respectively. The filament identification is not sensitive to these choices, one only requires the presence of density maxima at the ends of the filament. The density field is estimated using the Delaunay Tessellation Field Estimator (DTFE), which is implemented using the \texttt{delaunay\_3D} function of \disp. This DTFE density is smoothed using the \texttt{netconv} function with the parameter \texttt{nsmo=5}. The filament identification is done using the \texttt{mse} function, with \texttt{nsig=4}. Finally, the binary output is converted to human readable form using \texttt{skelconv}, with the keyword \texttt{-breakdown}, which merges overlapping filament segments.

Along with the spine of the filament, \disp\ also produces some nonphysical filaments, and the output needs to be cleaned to remove these artifacts. First, we remove any loops formed between two critical points that are not maxima. We also remove any saddle or bifurcation point that is linked to a single filament, and the filament associated with it. This is done recursively, until no freely hanging filament (one which ends at a point that is not a maximum) remains. Now, starting from one node, we loop over all the remaining filaments, joining those which share a common end point, until the other node is reached. This gives a single connected filament, which is used to estimate the profiles in section~\ref{sec:error on the spine}. Note that since here we have a single physical filament, this choice of the filament joining the two nodes is unique. However, in cosmological simulations, one might have multiple filaments joining at a point that is not a maximum (node). Then, the joining procedure is not so trivial, and one will have to make a choice (say choosing the longest or the shortest path joining two nodes).

\end{document}